\documentclass[aps,prc,twocolumn,groupedaddress]{revtex4}

\usepackage{graphicx}
\usepackage{amssymb}
\usepackage{amsmath}
\usepackage{color}
\newcommand{\leftsub}[2]{{\vphantom{#2}}_{#1}{#2}}
\begin{document}


\title{{\em Ab initio} many-body calculations of deuteron-$^4$He scattering and $^6$Li states}


\author{Petr Navr{\'a}til$^{1,2}$ and Sofia Quaglioni$^2$}
\affiliation{$^1$TRIUMF, 4004 Wesbrook Mall, Vancouver, BC, V6T 2A3, Canada\\
$^2$Lawrence Livermore National Laboratory, P.O. Box 808, L-414, Livermore, CA 94551, USA}

%
\date{\today}
\begin{abstract}
We extend the {\it ab initio} no-core shell model/resonating-group method (NCSM/RGM) to 
projectile-target binary-cluster states where the projectile is a deuteron. We discuss the formalism in detail and give algebraic expressions for the integration kernels. Using a soft similarity-renormalization-group evolved chiral nucleon-nucleon potential, we calculate deuteron-$^4$He scattering and investigate $^6$Li bound and unbound states. 
Virtual three-body breakup effects are obtained in an approximated way by including excited pseudo-states of the deuteron in the calculation.
We compare our results to experiment and to a standard NCSM calculation for $^6$Li.
\end{abstract}


\maketitle

\section{Introduction}
\label{introduction}

{\it Ab initio} many-body calculations of nuclear scattering and reactions pose a challenge to nuclear theory. For $A=3$ and 4 nucleon systems, the Faddeev~\cite{Witala01} and Faddeev-Yakubovsky~\cite{Lazauskas05} as well
as the hyperspherical harmonics (HH) \cite{Pisa} or the Alt, Grassberger and Sandhas (AGS) 
\cite{Deltuva} methods are applicable and successful. For systems with more than four nucleons, only very few approaches presently exist among which the Green's function Monte Carlo method applied recently to the calculation of the $n$-$^4$He scattering~\cite{GFMC_nHe4}.

Recently we combined the resonating-group method (RGM)~\cite{RGM,RGM1,RGM2,RGM3,Lovas98,Hofmann08} and the {\em ab initio} no-core shell model (NCSM)~\cite{NCSMC12}, into a new many-body approach~\cite{NCSMRGM,NCSMRGM_PRC,NCSMRGM_IT} ({\em ab initio} NCSM/RGM) capable of treating bound and scattering states of light nuclei in a unified formalism, starting from the fundamental inter-nucleon interactions. 
So far, applications have been limited to the description of projectile-target scattering where the projectile is a single nucleon. In particular, we first studied the $n\,$-${}^3$H, $n\,$-${}^4$He, $p\,$-${}^{3,4}$He, $n\,$-${}^{10}$Be, scattering processes~\cite{NCSMRGM,NCSMRGM_PRC} and later also the $n$-$^7$Li, $p$-$^7$Be scattering as well as nucleon scattering on $^{12}$C and $^{16}$O~\cite{NCSMRGM_IT}.
 In the present paper we extend the formalism to the case of a two-nucleon projectile and perform calculations for 
 deuteron-$^4$He (or $d$-$\alpha$) scattering. Simultaneously, we investigate the $d$-$\alpha$ bound state and compare our results to a standard NCSM calculation for $^6$Li. It should be emphasized that the 
present formalism is general and applicable to any other target nucleus, i.e. 
to any deuteron-nucleus system. 

The deuteron is weakly bound and can be easily deformed. Its polarization and virtual breakup cannot be neglected even at very low energies. A proper treatment of 
these effects requires 
the inclusion of three-body continuum states: neutron-proton-nucleus. This is very challenging. Even though the extension of the RGM formalism to include three-body clusters is feasible~\cite{3bbound1,3bcont2}, in this first application we limit ourselves to two-body clusters only and approximate 
virtual three-body breakup effects by discretizing the continuum with excited deuteron pseudo-states.

Deuteron-$^4$He scattering was investigated within the binary-cluster RGM formalism in the past~\cite{Thompson73,Kanada82,Kanada85}. However, the present investigation is the first that uses accurate nucleon-nucleon ($NN$) interactions (i.e. such that fit the $NN$ phase shifts with high precision) and 
many-body cluster wave functions obtained consistently from the same Hamiltonian. We do not fit or adjust any parameters, rather we systematically investigate the convergence of our results with respect to the size of the harmonic oscillator (HO) basis used to expand the cluster wave functions and localized parts of the RGM integration kernels as well as with respect to the number of deuteron pseudo-states and/or 
$^4$He excited states included in the calculation. We compare our results to a standard {\it ab initio} NCSM calculation for $^6$Li 
that uses the same $NN$ potential. In this study, we employ a similarity renormalization group (SRG) ~\cite{SRG,Roth_SRG} evolved chiral N$^3$LO $NN$ potential~\cite{N3LO} (SRG-N$^3$LO) that is soft enough 
for us to reach convergence within about $12{-}14\hbar\Omega$ HO excitations in the basis expansion.

In Sect.~\ref{formalism}, we briefly overview the general features of the NCSM/RGM formalism and present for the first time 
algebraic expressions for the NCSM/RGM integration kernels when the projectile nucleus has mass number $a=2$.  
The matrix elements of the norm kernel are given in this section, while those of the Hamiltonian kernel are presented in Appendix~\ref{appA}. 
In Sect.~\ref{results}, we discuss our results  
for $d$-$\alpha$ scattering and bound-state calculations. We show the calculated phase shifts and cross sections and compare the deuteron-$^4$He results to 
$^6$Li {\it ab initio} NCSM calculations with the same Hamiltonian. Conclusions and outlook are given in Sect.~\ref{conclusions}.

\section{Formalism}
\label{formalism}
In the present paper we apply the NCSM/RGM formalism introduced in Ref.~\cite{NCSMRGM_PRC} to the description of deuteron-nucleus collisions. While the derivation of the integration kernels was specialized for projectile-target basis states with a single-nucleon projectile, the theoretical framework presented in Ref.~\cite{NCSMRGM_PRC} is general and fully applicable to the present case. In this section we briefly revisit the NCSM/RGM formalism and provide algebraic expressions of the integration kernels for the specific case of a two-nucleon projectile.

Following the notation of Ref.~\cite{NCSMRGM_PRC}, the wave function for a scattering process involving 
a two-nucleon projectile and a target nucleus
can be cast in the form 
\begin{equation}
|\Psi^{J^\pi T}\rangle = \sum_{\nu} \int dr \,r^2\frac{g^{J^\pi T}_\nu(r)}{r}\,\hat{\mathcal A}_{\nu}\,|\Phi^{J^\pi T}_{\nu r}\rangle\,, \label{trial}
\end{equation}
through an expansion over  binary-cluster channel-states of channel spin $s$, relative angular momentum $\ell$, total angular momentum $J$, parity $\pi$, and isospin $T$,
\begin{eqnarray}
|\Phi^{J^\pi T}_{\nu r}\rangle &=& \Big [ \big ( \left|A{-}2\, \alpha_1 I_1^{\,\pi_1} T_1\right\rangle \left |2\,\alpha_2 I_2^{\,\pi_2} T_2\right\rangle\big ) ^{(s T)}\nonumber\\
&&\times\,Y_{\ell}\left(\hat r_{A-2,2}\right)\Big ]^{(J^\pi T)}\,\frac{\delta(r-r_{A-2,2})}{rr_{A-2,2}}\,.\label{basis}
\end{eqnarray}
The above basis states are uniquely identified by the channel index $\nu=\{A{-}2\,\alpha_1I_1^{\,\pi_1} T_1;\, 2\, \alpha_2 I_2^{\,\pi_2} T_2;\, s\ell\}$.
The internal wave functions of the colliding nuclei contain
$A{-}2$ and $2$ nucleons ($A{>}2$), respectively, are antisymmetric under exchange of internal nucleons, and depend on translationally invariant internal coordinates. They are eigenstates of $H_{(A-2)}$ and $H_{(2)}$, the ($A{-}2$)- and two-nucleon intrinsic Hamiltonians ($I_i$, $\pi_i$, $T_i$ and $\alpha_i$ denote respectively spin, parity, isospin and additional quantum numbers of the $i$-th cluster). The clusters centers of mass are separated by the relative vector ($\vec{r}_i$ being the position vector of the $i$-th nucleon)
\begin{equation}
\vec r_{A-2,2} = r_{A-2,2}\hat r_{A-2,2}= \frac{1}{A - 2}\sum_{i = 1}^{A - 2} \vec r_i - \frac{1}{2}\sum_{j = A - 1}^{A} \vec r_j\,.
\end{equation}
In Eq.~(\ref{basis}), the residual anti-symmetrization for exchange of nucleons pertaining to different clusters is guaranteed by the
anti-symmetrizer for the $(A{-}2,2)$ partition 
\begin{eqnarray}
\label{antisym}
\hat{\mathcal{A}}_{\nu} &\equiv& \hat{\mathcal{A}}_{(A-2,2)}\nonumber\\
&=&  C\left[1-\sum_{i=1}^{A-2}\sum_{k=A-1}^A \hat P_{i,k} + \sum_{i<j=1}^{A-2}\hat P_{i,A-1} \hat P_{j,A} \right],
\end{eqnarray}
where $C$ is the normalization constant $\sqrt{\frac{2}{A(A-1)}}$\,.
The unknown relative-motion wave functions $g^{J^\pi T}_\nu(r)$
can be determined by solving the many-body Schr\"odinger equation in the Hilbert space spanned by the basis states $\hat{\mathcal A}_{\nu}\,|\Phi^{J^\pi T}_{\nu r}\rangle$:
\begin{equation}
\sum_{\nu}\int dr \,r^2\left[{\mathcal H}^{J^\pi T}_{\nu^\prime\nu}(r^\prime, r)-E\,{\mathcal N}^{J^\pi T}_{\nu^\prime\nu}(r^\prime,r)\right] \frac{g^{J^\pi T}_\nu(r)}{r} = 0\,,\label{RGMeq}
\end{equation}
where 
\begin{eqnarray}
{\mathcal H}^{J^\pi T}_{\nu^\prime\nu}(r^\prime, r) &=& \left\langle\Phi^{J^\pi T}_{\nu^\prime r^\prime}\right|\hat{\mathcal A}_{\nu^\prime}H\hat{\mathcal A}_{\nu}\left|\Phi^{J^\pi T}_{\nu r}\right\rangle\,,\label{H-kernel}\\
{\mathcal N}^{J^\pi T}_{\nu^\prime\nu}(r^\prime, r) &=& \left\langle\Phi^{J^\pi T}_{\nu^\prime r^\prime}\right|\hat{\mathcal A}_{\nu^\prime}\hat{\mathcal A}_{\nu}\left|\Phi^{J^\pi T}_{\nu r}\right\rangle\,,\label{N-kernel}
\end{eqnarray}
are the Hamiltonian and norm kernels, respectively.
Here $E$ is the 
total energy in the center-of-mass (c.m.) frame, and $H$ is the intrinsic $A$-nucleon microscopic Hamiltonian, for which it is useful to use the decomposition, e.g.:
\begin{equation}\label{Hamiltonian}
H=T_{\rm rel}(r)+ {\mathcal V}_{\rm rel} +\bar{V}_{C}(r)+H_{(A-2)}+H_{(2)}\,.
\end{equation}
Further, $T_{\rm rel}(r)$ is the relative kinetic energy 
and ${\mathcal V}_{\rm rel}$ is the sum of all interactions between nucleons belonging to different clusters after subtraction of the average Coulomb interaction between them, explicitly singled out in the term $\bar{V}_{C}(r)=Z_{1\nu}Z_{2\nu}e^2/r$, where $Z_{1\nu}$ and $Z_{2\nu}$ are the charge numbers of the clusters in channel $\nu$: 
\begin{equation}
{\mathcal V}_{\rm rel} = \sum_{i=1}^{A-2}\sum_{j=A-1}^AV_{ij}-\bar{V}_{C}(r) \label{pot}\,.
\end{equation}
The $V_{ij}$ interaction consists of the strong and Coulomb part. Thanks to the subtraction of $\bar{V}_{C}(r)$, the overall Coulomb contribution presents a $r^{-2}$ behavior, as the distance $r$ between the two clusters increases. Therefore, ${\mathcal V}_{\rm rel}$ is localized also in presence of the Coulomb force. In this paper, we limit our calculations to the use of  a two-nucleon interaction only, but the formalism can be generalized to include the three-nucleon interaction in a straightforward way.

\subsection{Norm kernel}

For definitions and details regarding the derivations outlined in this and the next section we refer the interested reader to Secs.\ II.C.1 and II.C.2 of Ref.~\cite{NCSMRGM_PRC}.

Because the wave functions of both (A-2)-nucleon and two-nucleon clusters are anti-symmetric under exchange of internal nucleons, the norm kernel~(\ref{N-kernel}) for the same, $(A{-}2,2)$, mass partition in  both the initial and final state can be written as
\begin{figure}[h]
\includegraphics*[width=0.8\columnwidth]{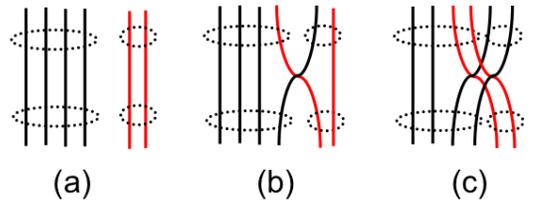}
\caption{(Color online) Diagrammatic representation of: ($a$) ``direct", ($b$) ``one-nucleon-exchange" and ($c$) ``two-nucleon-exchange"  components of the norm kernel. The groups of circled lines represent the $(A{-}2)$- and two-nucleon clusters. Bottom and upper part of the diagram represent initial and final states, respectively.}\label{diagram-norm}
\end{figure}
\begin{widetext}
\begin{eqnarray}
\label{Am22norm}
\mathcal{N}_{\nu'\nu}^{J^\pi T}(r', r) &= &
\left<\Phi_{\nu' r'}^{J^\pi T}\right|\hat{\mathcal{A}}^2_{(A-2,2)}\left|\Phi_{\nu r}^{J^\pi T}\right>\\
& = & \delta_{\nu^\prime \nu}\frac{\delta(r^\prime - r)}{r^\prime r} + \sum_{n^\prime n} R_{n^\prime \ell^\prime} (r^\prime) R_{n \ell} (r) 
\left[ -2(A-2) \left<\Phi_{\nu' n'}^{J^\pi T}\right| \hat P_{A-2,A}\left|\Phi_{\nu n}^{J^\pi T}\right>  \right.
\nonumber\\
&&\;\;\phantom{ \delta_{\nu^\prime \nu}\frac{\delta(r^\prime - r)}{r^\prime r} + \sum_{n^\prime n} R_{n^\prime \ell^\prime} (r^\prime) R_{n \ell} (r)}
\left. + \frac {(A-2)(A-3)}{2} \left<\Phi_{\nu' n'}^{J^\pi T}\right| \hat P_{A-2,A} P_{A-3,A-1}\left|\Phi_{\nu n}^{J^\pi T}\right> \right]\,,
\end{eqnarray}
\end{widetext}
where $\left|\Phi_{\nu n}^{J^\pi T}\right> $ is the translationally-invariant HO channel state introduced in Eq.~(22) of Ref.~\cite{NCSMRGM_PRC}, here for clarity:
\begin{eqnarray}
|\Phi^{J^\pi T}_{\nu n}\rangle &=& \Big [ \big ( \left|A{-}2\, \alpha_1 I_1^{\,\pi_1} T_1\right\rangle \left |2\,\alpha_2 I_2^{\,\pi_2} T_2\right\rangle\big ) ^{(s T)}\nonumber\\
&&\times\,Y_{\ell}\left(\hat r_{A-2,2}\right)\Big ]^{(J^\pi T)}\,R_{n \ell}(r_{A-2,2})\,.\label{basis-n}
\end{eqnarray}
Three terms contribute to the norm kernel~(\ref{Am22norm}):
A direct term, in which initial and final states are identical (corresponding to diagram $(a)$ of Fig.~\ref{diagram-norm}); 
a one-nucleon exchange term, corresponding to diagram $(b)$ of Fig.~\ref{diagram-norm}; and, finally, a two-nucleon exchange term, corresponding to diagram $(c)$ of Fig.~\ref{diagram-norm}.

In this paper, the localized parts of the integration kernels~(\ref{H-kernel}) and~(\ref{N-kernel}) are obtained in two steps. First,  matrix elements of translationally-invariant operators (for the norm $\hat P_{A-2,A}$ and $\hat P_{A-2,A} \hat P_{A-3,A-1}$) are calculated in the Slater-determinant (SD) basis, in which the eigenstates of the $(A{-}2)$-nucleon fragment are expanded in HO Slater determinants: 
\begin{eqnarray}
|\Phi^{J^\pi T}_{\nu n}\rangle_{\rm SD}   &=&    \Big [\big (\left|A{-}2\, \alpha_1 I_1 T_1\right\rangle_{\rm SD} 
\left |2\,\alpha_2 I_2 T_2\right\rangle\big )^{(s T)}\nonumber\\
&&\times Y_{\ell}(\hat R^{(2)}_{\rm c.m.})\Big ]^{(J^\pi T)} R_{n\ell}(R^{(2)}_{\rm c.m.})\,.
\label{SD-basis}
\end{eqnarray}
Second, the corresponding translationally-invariant matrix elements on the basis (\ref{basis-n}) are  recovered through a transformation as described in Sect.\ II.C.2 of Ref.~\cite{NCSMRGM_PRC}, Eq.~(32).
Here we remind that 
the eigenstates of the $(A{-}2)$-nucleon fragment in the SD basis, 
\begin{equation}
\langle\vec r_1\!\cdots\vec r_{A-2}\sigma_1\!\cdots\sigma_{A-2}\tau_1\!\cdots\tau_{A-2}|A{-}2\,\alpha_1I^{\pi_1}_1T_1\rangle_{\rm SD},
\end{equation}
are related to the transaltionally-invariant eigenstates by the expression 
\begin{equation}
\left|A{-}2\, \alpha_1 I_1 T_1\right\rangle_{\rm SD} = \left|A{-}2\, \alpha_1 I_1 T_1\right\rangle\,\varphi_{00}(\vec R^{(A-2)}_{\rm c.m.})\,,
\label{SD-eigenstate}
\end{equation}  
and the c.m. coordinates introduced in Eqs.~(\ref{SD-basis}) and (\ref{SD-eigenstate}) are given by:
\begin{equation}
\vec R^{(A - 2)}_{\rm c. m.} = \sqrt{\frac{1}{A - 2}}\sum_{i = 1}^{A - 2} \vec r_i\;; \quad \vec R^{(2)}_{\rm c. m.} = \sqrt{\frac{1}{2}}\sum_{i = A - 1}^{A} \vec r_i\,.
\end{equation}
The calculation of matrix elements 
in the basis (\ref{SD-basis}) is most efficiently achieved by first performing a transformation to a new SD basis:
\begin{eqnarray}
|\Phi^{J^\pi T}_{\nu n}\rangle_{\rm SD}   &=&  \sum 
\left\{ \begin{array}{@{\!~}c@{\!~}c@{\!~}c@{\!~}} 
I_1 & I_2 & s \\[2mm] 
\ell & J & j 
\end{array}\right\}   
\left\{ \begin{array}{@{\!~}c@{\!~}c@{\!~}c@{\!~}} 
\ell & L_{ab} & {\ell}_2 \\[2mm] 
 s_2 & I_2 & I 
\end{array}\right\}
\left\{ \begin{array}{@{\!~}c@{\!~}c@{\!~}c@{\!~}} 
{\ell}_a & {\ell}_b & L_{ab} \\[2mm] 
 \frac{1}{2} & \frac{1}{2} & s_2 \\[2mm] 
 j_a & j_b & I 
\end{array}\right\}  
\nonumber\\[2mm] 
&&\times 
(-1)^{I_1+J+\ell+\ell_2+T_2} \, \hat{s}\, \hat{I}\, \hat{I}_2\, \hat{s}_2 \, \hat{j}_a\, \hat{j}_b\, \hat{L}_{ab}^2
\nonumber\\[2mm]
&&\times
 \left\langle n_a {\ell}_a n_b {\ell}_b L_{ab} \left. \right| n \ell n_2 {\ell}_2 L_{ab}\right\rangle_{d=1}
\nonumber \\[2mm] 
&& \times
\left \langle n_2 {\ell}_2 s_2 I_2 T_2 \left. \right| 2\,\alpha_2 I_2 T_2\right\rangle \; |\Phi^{J^\pi T}_{\kappa_{ab}}\rangle_{\rm SD}
\label{SD-basis-SNP}
\end{eqnarray}
where the sum runs over the quantum numbers $n_2,\ell_2, s_2$, $n_a, \ell_a, j_a$, $n_b, \ell_b, j_b$, $L_{ab}$, and $I$,
$ \langle n_2 \ell_2 s_2 I_2 T_2 | 2\,\alpha_2 I_2 T_2\rangle $ is the projectile wave function expanded in the relative-coordinate HO basis, $\hat{s}=\sqrt{2s{+}1}$ etc., and $\langle n_a {\ell}_a n_b {\ell}_b L_{ab} | n \ell n_2 {\ell}_2 L_{ab}\rangle_{d=1}$ indicates an HO bracket for two particles with identical masses. In addition, we introduced  the cumulative quantum number $\kappa_{ab} \equiv \{ A{-}2 \, \alpha_1 I_1 T_1$; $n_a \ell_a j_a \tfrac12; n_b \ell_b j_b \tfrac12; I T_2\}$ and the new SD channel states:
\begin{align}
|\Phi^{J^\pi T}_{\kappa_{ab}}\rangle_{\rm SD} &= \Big [\left|A{-}2\, \alpha_1 I_1 T_1\right\rangle_{\rm SD} 
\left(\varphi_{n_a \ell_a j_a \frac12} (\vec{r}_A \sigma_A \tau_A) \right.
\nonumber \\
&\quad\phantom{=}\times \left. \varphi_{n_b \ell_b j_b \frac12} (\vec{r}_{A-1} \sigma_{A-1} \tau_{A-1})\right)^{(I T_2)}\Big ]^{(J^\pi T)}.
\label{SD-basis-ab}
\end{align}
Using 
the basis states of Eq.~(\ref{SD-basis-ab}) to evaluate the matrix elements of the transposition operators appearing in 
Eq.~(\ref{Am22norm}) results in the following expressions:
\begin{widetext}
\begin{align}
\leftsub{\rm SD}{\left\langle\Phi^{J^\pi T}_{\kappa_{ab}^\prime}\left| \hat{P}_{A-2,A} \right|\Phi^{J^\pi T}_{\kappa_{ab}}\right\rangle}_{\rm SD}  &=
\delta_{b, b^\prime} \frac{1}{A-2}\sum_{K \tau}
\left\{ \begin{array}{@{\!~}c@{\!~}c@{\!~}c@{\!~}} 
I_1 & K & I_1^\prime \\[2mm] 
I^\prime & J & I 
\end{array}\right\}
\left\{ \begin{array}{@{\!~}c@{\!~}c@{\!~}c@{\!~}} 
j_a & j_a^\prime & K \\[2mm] 
I^\prime & I & j_b 
\end{array}\right\}
\left\{ \begin{array}{@{\!~}c@{\!~}c@{\!~}c@{\!~}} 
T_1 & \tau & T_1^\prime \\[2mm] 
T_2^\prime & T & T_2 
\end{array}\right\}
\left\{ \begin{array}{@{\!~}c@{\!~}c@{\!~}c@{\!~}} 
\frac{1}{2} & \frac{1}{2} & \tau \\[2mm] 
T_2^\prime & T_2 & \frac{1}{2} 
\end{array}\right\}
\nonumber \\[2mm]
& \phantom{=}\times 
 (-1)^{I+I^\prime-I_1-J+j_b-j_a+K} (-1)^{T_2+T_2^\prime-T_1-T+\tau}\, \hat{I}\, \hat{I}^\prime\, \hat{K}\,
\hat{T}_2 \, \hat{T}_2^\prime \, \hat{\tau}
\nonumber \\[2mm]
& \phantom{=}\times 
\leftsub{\rm SD}{\left\langle A{-}2\, \alpha_1^\prime I_1^\prime T_1^\prime \left | \left | \left | 
\left(a^\dagger_{a}\tilde{a}^{\phantom l}_{a^\prime}\right)^{(K\tau)}\right | \right | \right |A{-}2\, \alpha_1 I_1 T_1\right \rangle}_{\rm SD} \,,
\label{P_AAm2_SD-basis-ab}
\end{align}
%
and
%
\begin{align}
&\leftsub{\rm SD}{\left\langle\Phi^{J^\pi T}_{\kappa_{ab}^\prime}\left| \hat{P}_{A-2,A}\hat{P}_{A-3,A-1} \right|\Phi^{J^\pi T}_{\kappa_{ab}}\right\rangle}_{\rm SD}  
\nonumber\\[2mm]
&\qquad= 
\frac{1}{(A-2)(A-3)}{\sum_{K \tau}}
\left\{ \begin{array}{@{\!~}c@{\!~}c@{\!~}c@{\!~}} 
I_1 & K & I_1^\prime \\[2mm] 
I^\prime & J & I 
\end{array}\right\}
\left\{ \begin{array}{@{\!~}c@{\!~}c@{\!~}c@{\!~}} 
T_1 & \tau & T_1^\prime \\[2mm] 
T_2^\prime & T & T_2 
\end{array}\right\}
\, (-1)^{I_1+I+I^\prime+J+j_a^\prime+j_b^\prime} (-1)^{T_1+T_2+T_2^\prime+T+1}  \, \hat{K}\, \hat{\tau}
\nonumber\\[2mm]
&\qquad\phantom{=} \times  
\leftsub{\rm SD}{\left\langle A{-}2\, \alpha_1^\prime I_1^\prime T_1^\prime \left |\left |\left | 
\left((a^\dagger_a a^\dagger_b)^{(I T_2)}(\tilde{a}_{b^\prime}\tilde{a}_{a^\prime})^{(I^\prime T_2^\prime)}\right)^{(K\tau)} \right |\right |\right |A{-}2\, \alpha_1 I_1 T_1\right \rangle}_{\rm SD} 
\label{P_AAm2_P_Am3Am1_SD-basis-ab}
\end{align}
\end{widetext}
where the indexes $a$ and $b$ represent the sets of single-particle quantum numbers  $\{n_a \ell_a j_a \frac12 \}$ and $\{n_b \ell_b j_b \frac12 \}$, respectively, such that $a^\dagger_a\equiv a^\dagger_{n_a\ell_aj_a \frac12}$ etc., $a^\prime$ and $b^\prime$ are analogous indexes associated with the primed quantum numbers, $\kappa_{ab}^\prime = \{ A{-}2 \, \alpha^\prime_1 I^\prime_1 T^\prime_1; a^\prime; b^\prime; I^\prime T^\prime_2\}$, and, finally, $\tilde{a}_{n \ell j m\frac{1}{2}m_t}=(-1)^{j-m+\frac{1}{2}-m_t}a_{n \ell j-m\frac{1}{2}-m_t}$. In addition, we note that  Eqs.~(\ref{P_AAm2_SD-basis-ab}) and (\ref{P_AAm2_P_Am3Am1_SD-basis-ab}) depend on the one- and two-body density matrix elements (OBDME and TBDME), respectively, of the target nucleus. 

\subsection{Hamiltonian Kernel}

\begin{figure}
\includegraphics*[width=0.85\columnwidth]{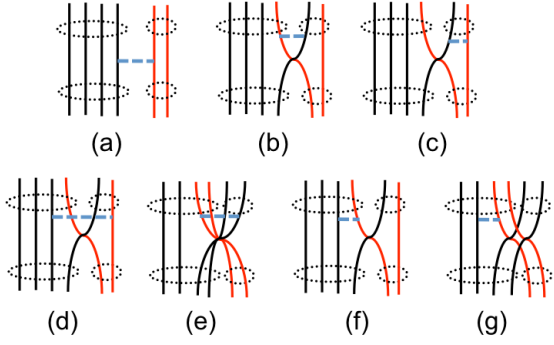}
\caption{(Color online) Diagrammatic representation of: ``direct potential",  ($a$)-($b$), and ``exchange-potential", ($c$)-($g$), components of the potential kernel. The groups of circled lines represent the $(A{-}2)$- and two-nucleon clusters. Bottom and upper part of the diagram represent initial and final states, respectively.}\label{diagram-potential}
\end{figure}

The Hamiltonian kernel (\ref{H-kernel}) for the same, $(A{-}2,2)$, mass partition in both the initial and final state 
can be cast in the form
\begin{widetext}
\begin{eqnarray}
\mathcal{H}_{\nu'\nu}^{J^\pi T}(r',r)&=&
\left<\Phi_{\nu' r'}^{J^\pi T}\right|\hat{\mathcal{A}}_{(A-2,2)}H\hat{\mathcal{A}}_{(A-2,2)}
\left|\Phi_{\nu r}^{J^\pi T}\right>\notag
 =\left<\Phi_{\nu' r'}^{J^\pi T}\right|H\hat{\mathcal{A}}^2_{(A-2,2)}
\left|\Phi_{\nu r}^{J^\pi T}\right>\notag\\
&=&\left[{T}_{\rm rel}(r')+\bar{V}_C(r')+E_{\alpha_1'}^{I_1'T_1'} +E_{\alpha_2'}^{I_2'T_2'}\right]
\mathcal{N}_{\nu'\nu}^{J^\pi T}(r', r)+\mathcal{V}^{J^\pi T}_{\nu' \nu}(r',r),
\end{eqnarray}
where the potential kernel is defined by
\begin{align}
\mathcal{V}^{J^\pi T}_{\nu' \nu}(r',r) &= 
\left<\Phi_{\nu' r'}^{J^\pi T}\right|\mathcal{V}_{\rm rel}\hat{\mathcal{A}}^2_{(A-2,2)} \left|\Phi_{\nu r}^{J^\pi T}\right> 
\label{V-kernel}\\
& =  \sum_{n^\prime n} R_{n^\prime \ell^\prime} (r^\prime) R_{n \ell} (r)\,  \left[ 2(A-2) \left<\Phi_{\nu' n'}^{J^\pi T}\right| V_{A-2,A-1}(1-\hat P_{A-2,A-1})\left|\Phi_{\nu n}^{J^\pi T}\right>  \right. \nonumber\\
& \quad\phantom{ =\sum_{n^\prime n} R_{n^\prime \ell^\prime} (r^\prime) R_{n \ell} (r)}
- 2(A-2) \left<\Phi_{\nu' n'}^{J^\pi T}\right| V_{A-2,A}\hat P_{A-2,A-1}\left|\Phi_{\nu n}^{J^\pi T}\right>  \nonumber\\
& \quad\phantom{ =\sum_{n^\prime n} R_{n^\prime \ell^\prime} (r^\prime) R_{n \ell} (r)}  
- 2(A-2)(A-3) \left<\Phi_{\nu' n'}^{J^\pi T}\right| V_{A-3,A}(1-\hat P_{A-3,A}) \hat P_{A-2,A-1}\left|\Phi_{\nu n}^{J^\pi T}\right> \nonumber\\
&\quad\phantom{ =\sum_{n^\prime n} R_{n^\prime \ell^\prime} (r^\prime) R_{n \ell} (r)}
- 2(A-2)(A-3) \left<\Phi_{\nu' n'}^{J^\pi T}\right| V_{A-3,A-1}\hat P_{A-2,A-1}\left|\Phi_{\nu n}^{J^\pi T}\right> \nonumber\\
&\quad\phantom{ =\sum_{n^\prime n} R_{n^\prime \ell^\prime} (r^\prime) R_{n \ell} (r)}
\left. + (A-2)(A-3)(A-4) \left<\Phi_{\nu' n'}^{J^\pi T}\right| V_{A,A-4}(1-\hat P_{A-2,A-1}) \hat P_{A-3,A}\left|\Phi_{\nu n}^{J^\pi T}\right> \right]\,,
\label{antisym_HAm2}
\end{align}
\end{widetext}
Clearly, for the ($A{-}2$,$2$) partition the potential kernel presents a much more complicated expression than the norm kernel.
We identify five separate terms
corresponding to the nine diagrams presented in Fig.~\ref{diagram-potential}. The  first ``direct-potential'' term on the right-hand side (rhs) of Eq.~(\ref{antisym_HAm2}) corresponds to diagrams (a) and (b), the second term corresponds to diagram (c), while diagrams (d) and (e) represent the third term. The last two terms are then depicted schematically by diagrams (f) and (g), respectively.
Matrix elements of each of these terms in the basis (\ref{SD-basis-ab}) are given in Appendix~\ref{appA}. The first two terms, (\ref{V-kernel_1}) and (\ref{V-kernel_2}), depend on the OBDME of the target nucleus; the second two terms, (\ref{V-kernel_3}) and (\ref{V-kernel_4}), depend on the TBDME of the target nucleus; and, finally, the last term, (\ref{V-kernel_5}), depends on the three-body density of the target nucleus. The three-body density matrix elements can be obviously re-coupled in different ways. Here, we selected a particular angular-momentum coupling that results in the simplest expression for our Hamiltonian kernel matrix element. In general, it is a challenge to compute three-body density matrix elements, in particular due to their rapidly increasing number in the multi-major shell basis spaces. However, since in the present paper we focus on the $A=6$, $d$-$^4$He and $^6$Li systems, we can take advantage of the completeness of the $(A{-}5)$-body eigenstates and re-write the expression (\ref{V-kernel_5}) in the form
\begin{widetext}
\begin{align}
 &\leftsub{\rm SD}{\left\langle\Phi^{J^\pi T}_{\kappa_{ab}^\prime}\left|V_{A,A-4} \hat{P}_{A-2,A-1}\hat{P}_{A-3,A} \right|\Phi^{J^\pi T}_{\kappa_{ab}}\right\rangle}_{\rm  SD} 
 \nonumber \\[2mm]
 & \qquad =
\frac{1}{2}\frac{1}{(A-2)(A-3)(A-4)}
\sum
\left\{ \begin{array}{@{\!~}c@{\!~}c@{\!~}c@{\!~}} 
J_{de} & j_b^\prime & X \\[2mm] 
I^\prime & j_e^\prime & j_a^\prime 
\end{array}\right\}
\left\{ \begin{array}{@{\!~}c@{\!~}c@{\!~}c@{\!~}} 
T_{de} & \frac{1}{2} & \tau_X \\[2mm] 
T_2^\prime & \frac{1}{2} & \frac{1}{2} 
\end{array}\right\}
\left\{ \begin{array}{@{\!~}c@{\!~}c@{\!~}c@{\!~}} 
I_1 & I & J \\[2mm] 
X & j_e^\prime & I^\prime \\[2mm] 
I_\beta & Y & I_1^\prime 
\end{array}\right\}
\left\{ \begin{array}{@{\!~}c@{\!~}c@{\!~}c@{\!~}} 
T_1 & T_2 & T \\[2mm] 
\tau_X & \frac{1}{2} & T_2^\prime \\[2mm] 
T_\beta & \tau_Y & T_1^\prime 
\end{array}\right\}
\nonumber \\[2mm]
& \qquad \phantom{=} \times
\hat{I}^\prime \, \hat{X} \, \hat{Y} \, \hat{J}_{de} \, \hat{T}_2^\prime \, \hat{\tau}_X \, \hat{\tau}_Y \, \hat{T}_{de} \,
(-1)^{I^\prime+J_{de}+J-I_1^\prime+j_e-j_d} \; (-1)^{T_2^\prime+T_{de}+T-T_1^\prime}
\nonumber \\[2mm]
&\qquad \phantom{=} \times
 \leftsub{\rm SD}{\left\langle A{-}2\, \alpha_1^\prime I_1^\prime T_1^\prime \left | \left |\left | 
\left((a^\dagger_a a^\dagger_b)^{(I T_2)} a^\dagger_{e^\prime}\right)^{(Y \tau_Y)}\right | \right | \right | A{-}5 \,\beta I_\beta T_\beta\right\rangle}_{\rm SD}
\nonumber \\[2mm]
&\qquad \phantom{=} \times
 \leftsub{\rm SD}{\left\langle  A{-}5 \,\beta I_\beta T_\beta \left |\left |\left | \left(\tilde{a}_{b^\prime} (\tilde{a}_{e}\tilde{a}_{d})^{(J_{de} T_{de})}\right)^{(X\tau_x)}
\right | \right | \right | A{-}2\, \alpha_1 I_1 T_1\right\rangle}_{\rm SD}
\nonumber \\[2mm]
&\qquad \phantom{=} \times
\sqrt{1+\delta_{a^\prime,e^\prime}} \, \sqrt{1+\delta_{d,e}}\;\;
\left\langle a^\prime e^\prime J_{de} T_{de} \left| V \right| d \,e\, J_{de} T_{de} \right\rangle 
\,,
\label{V-kernel_5-compl}
\end{align}
\end{widetext}
where the sum runs over the quantum numbers $\beta, I_\beta, T_\beta$, $d \equiv n_d \ell_d j_d \tfrac12$ etc., $e, e^\prime$, $J_{de}, T_{de}$, $X, Y, \tau_X$, and $\tau_Y$. 
For the present case of $A=6$, the states $|A{-}5 \,\beta I_\beta T_\beta\rangle_{\rm SD}$ reduce to the HO single particle states $|n_{\beta} l_{\beta} j_{\beta} \frac{1}{2}\rangle$ and the reduced matrix elements in Eq.~(\ref{V-kernel_5-compl}) that involve $^4$He eigenstates are straightforward to calculate.

We also note that the terms (\ref{V-kernel_1}) and (\ref{V-kernel_3}) are symmetric, while the remaining ones, (\ref{V-kernel_2}), (\ref{V-kernel_4}) and (\ref{V-kernel_5}), are not. Therefore, we introduce a Hermitized NCSM/RGM Hamiltonian, as discussed in detail in Ref.~\cite{NCSMRGM_PRC}, using $\hat{\mathcal A}H\hat{\mathcal A}=\frac12(\hat{\mathcal A}^2H+H\hat{\mathcal A}^2)$ (see Eq.~(42) in Ref.~\cite{NCSMRGM_PRC}).

\section{Application to the deuteron-$^4$He system}
\label{results}

The deuteron-nucleus formalism presented in the previous section is completely general. The simplest system to which it can be applied is deuteron-$^4$He for two reasons. First, the complicated calculation of the target three-body density needed to compute the last term on the right-hand side of Eq.~(\ref{antisym_HAm2}), given by Eq.~(\ref{V-kernel_5}), becomes straightforward for for $A=5$ and 6 (that is for $^3$H, $^3$He and $^4$He targets) using the completeness of the $(A{-}5)$-nucleon eigenstates as demonstrated in Eq.~(\ref{V-kernel_5-compl}). 
Second, the $^4$He nucleus is tightly bound with its first excited state at $E_x\approx 20$~MeV. A solution of the coupled-channel equations (\ref{RGMeq}) obtained limiting the target-states to the ground state (g.s.) is already a very good approximation for the $d$-$\alpha$ system.

To test the formalism, we use a soft SRG-N$^3$LO $NN$ potential with evolution parameter $\Lambda=1.5$~fm$^{-1}$. With this low value of $\Lambda$, 
our calculations reach convergence at $N_{\rm max}\approx 12$. We note, however, that a somewhat higher value of $\Lambda$ would result in a better agreement with experimental data as discussed later.
We benchmark our $d$-$\alpha$ NCSM/RGM  
results with standard NCSM calculations for $^6$Li. Any differences can then be attributed to missing degrees of freedom rather than to the model space truncation. 

Our calculation starts with the NCSM diagonalization of the Hamiltonian in the $N_{\rm max}\hbar\Omega$ HO basis for $d$ and $^4$He. Obtained eigenenergies and eigenfunctions serve then as input in Eq.~(\ref{RGMeq}). First, one-, two-, and three-body densities are calculated from the $^4$He wave functions, then the integration kernels are calculated. The localized parts of the integration kernels are expanded in the same $N_{\rm max}$ (or $N_{\rm max}+1$ depending on parity) HO basis space as the cluster eigenstates. The same HO frequency is used in all calculations. The wave functions of the $d$-$\alpha$ relative motion are found by solving (\ref{RGMeq}) with either bound-state or scattering-state boundary conditions 
by means of the microscopic R-matrix method on a Lagrange mesh~\cite{R-matrix} (with additional details given in Ref.~\cite{NCSMRGM_PRC}, Sect. II. F). 

\subsection{Bound-state calculations}

\begin{figure}
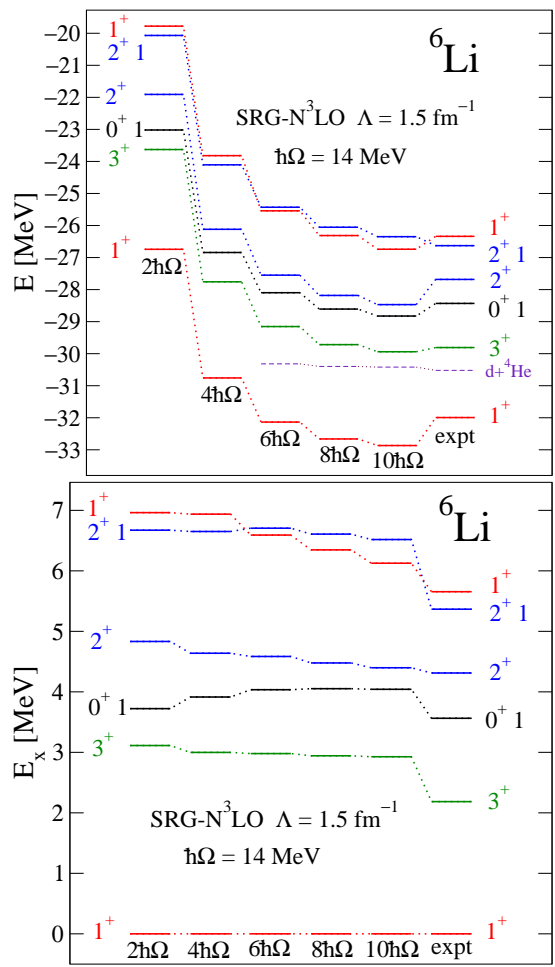

\begin{minipage}{8cm}
\includegraphics*[width=0.9\columnwidth]{Li6_srg-n3lo_1p5_14_spectra_vs_Nmax.eps}
\end{minipage}
\begin{minipage}{8cm}
\includegraphics*[width=0.9\columnwidth]{Li6_srg-n3lo_1p5_14_spectra_vs_Nmax_excited.eps}
\end{minipage}
\caption{(Color online) Absolute (top panel) and excitation (bottom panel) energies of $^6$Li calculated within the NCSM compared to experiment. The dashed lines in the top panel indicate the calculated NCSM ($N_{\rm max}=8,10,12$) and experimental $d$-$\alpha$ thresholds. The SRG-N$^3$LO $NN$ potential with $\Lambda=1.5$~ fm$^{-1}$ and the HO frequency of $\hbar\Omega=14$ MeV were used.}\label{Li6_NCSM}
\end{figure}

Our results for the ground states of deuteron, $^4$He and $^6$Li are presented in Table~\ref{tab:Egs_2-6}.
\begin{table}[tb]
\begin{ruledtabular}
\begin{tabular}{ccccc}
$E_{\rm g.s.}$ [MeV]   & $^{2}$H    & $^{4}$He    & $^{6}$Li (NCSM/RGM)  &$^{6}$Li (NCSM) \\
\hline
Calc.                        & -2.20     & -28.22      & -32.25      & -32.87  \\       
Expt.                        & -2.22     & -28.30      & -31.99      & -31.99  \\
\end{tabular}
\end{ruledtabular}
\caption{Calculated g.s.\ energies of $^2$H, $^4$He, and $^6$Li obtained by using the SRG-N$^3$LO $NN$ potential with $\Lambda=1.5$ fm$^{-1}$ are compared to the corresponding experimental values. The NCSM calculations for $^{2}$H, $^4$He and $^{6}$Li were performed in $N_{\rm max}=12,12$ and $10$ basis space, respectively.  The NCSM/RGM calculation included $^4$He and $^2$H ground states and 7 deuteron pseudo-states in each of the  $^3S_1$-$^3D_1$ and $^3D_3$-$^3G_3$ channels, as well as 5 pseudo-states in the $^3D_2$ channel. The HO frequency of $\hbar\Omega=14$ MeV was used.}\label{tab:Egs_2-6}
\end{table}
The convergence of the NCSM calculations can be judged from Fig.~\ref{Li6_NCSM}, where we show both absolute and excitation energies of $^6$Li as well as the $d$+$\alpha$ threshold. The $^4$He convergence is excellent and that of $^6$Li very good. The $^6$Li excited states are resonances, but within the NCSM calculation they are approximated by eigenstates expanded in the HO basis.
The $^4$He is slightly under-bound while the $^6$Li is over-bound by about 0.9 MeV due to the choice of a low $\Lambda$ value and the neglect of the SRG-induced three-body interaction~\cite{Li6_SRG}. We note that in the $NN$-only calculations of Ref.~\cite{Li6_SRG}, selecting $\Lambda\approx 2$~fm$^{-1}$ results in 
$^4$He and $^6$Li  binding energies closer to experiment. The excited states are correctly ordered except for the reversal of the $2^+ 1$ and the $1^+_2 0$ states. The splitting of the $2^+ 0$ and the $3^+ 0$ states is underestimated, a sign of weak spin-orbit interaction, most likely due to the neglect of the initial chiral three-nucleon interaction. One more feature to notice is the drop of the $2^+ 0$ and $1^+_2 0$ excitation energies with increasing $N_{\rm max}$. This is a consequence of the fact that these states are broader resonances compared to the $3^+ 0$ or the $T=1$ states.

\begin{figure}
\includegraphics*[width=0.9\columnwidth]{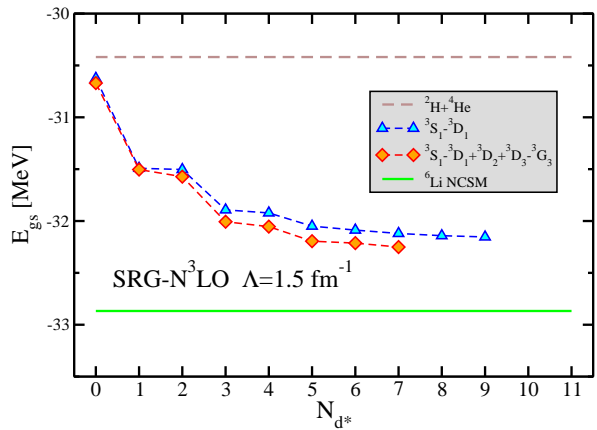}
\caption{(Color online) Dependence of the $^6$Li g.s.\ energy on the number of deuteron pseudo-states $N_{\rm d^*}$ in the $^3S_1$-$^3D_1$, $^3D_2$ and $^3D_3$-$^3G_3$ channels included in the calculation. Diamond symbols: All three channels included. Triangle symbols: Only $^3S_1$-$^3D_1$ channel included. The NCSM results (solid line) and the $^2$H+$^4$He threshold energy (dashed line) are also shown. Details of the calculations are described in the caption of Table~\protect\ref{tab:Egs_2-6}.\label{gs_en_on_pseudost}}
\end{figure}
The variation with respect to $N_{\rm max}$ of the $^6$Li g.s.\ energy calculated within the NCSM/RGM is similar to that of the $^4$He. The $N_{\rm max}=12$ model-space is sufficient to reach convergence. However, 
the deuteron is weakly bound: Its polarization and virtual breakup cannot be neglected. This is demonstrated in Fig.~\ref{gs_en_on_pseudost}. The NCSM/RGM calculation limited to the deuteron ground state binds $^6$Li by only about 200 keV contrary to the NCSM result of 2.4 MeV using the same Hamiltonian. To include deuteron polarization and virtual breakup properly one would have to extend the NCSM/RGM formalism to a three-cluster basis: $n$-$p$-$\alpha$. This is quite challenging. In the present work, we discretize the continuum by including excited deuteron pseudo-states in the NCSM/RGM coupled channel equations. The pseudo-states are obtained in the NCSM diagonalization. In Table~\ref{tab:pseudo-states}, we present the pseudo-state energies obtained in the $N_{\rm max}=12$ basis. The $^6$Li g.s.\ convergence with respect to the number of $d$ pseudo-states $N_{\rm d^*}$ included in the calculation is shown in Fig.~\ref{gs_en_on_pseudost}. The $S$-wave dominated (odd $N_{\rm d^*}$) pseudo-states in the $^3S_1$-$^3D_1$ channel have a quite dramatic influence on the $^6$Li binding energy. The  pseudo-states in the $^3D_2$ and $^3D_3$-$^3G_3$ channels are less important for the ground state but have a significant effect on the $2^+$ and the $3^+$ $^6${Li} resonances. By including 7 or 9 pseudo-states, we reach convergence with respect to the number of $d^*$ in the channels considered here (for the g.s.\ energy, the $N_{\rm d^*}=7$ and 9 results are within 30 keV of eachother). Still, the $^6$Li NCSM calculation contains more correlations as it produces a lower g.s.\ energy by about 600 keV or ~2\%. This can be seen from Table~\ref{tab:Egs_2-6} and Fig.~\ref{gs_en_on_pseudost}. The missing correlations in the NCSM/RGM calculation most likely include excitations of the $^4$He (of which we have included here only the ground state), as well as 
deuteron excitations in other channels.
\begin{table}[tb]
\begin{ruledtabular}
\begin{tabular}{cccc}
$E$ [MeV]   &  $^3S_1$-$^3D_1$    &   $^3D_2$  & $^3D_3$-$^3G_3$ \\
\hline
g.s.             &    -2.20     &    -       & - \\
$1^*$         &      4.50     &  7.53     & 7.61 \\ 
$2^*$         &      7.69     &  18.81   & 15.72 \\
$3^*$         &    15.20     &  35.05   &  19.32 \\
$4^*$         &    19.74     &  57.28   &  33.13 \\
$5^*$         &    31.90     &  87.88   &  36.47 \\
$6^*$         &    37.60     &   -         &  57.01 \\
$7^*$         &    55.95     &  -          &  60.13
\end{tabular}
\end{ruledtabular}
\caption{Calculated $^2$H g.s.\ and pseudo-state energies obtained using the SRG-N$^3$LO $NN$ potential with $\Lambda=1.5$ fm$^{-1}$ in the $N_{\rm max}=12$ basis space and HO frequency of $\hbar\Omega=14$~ MeV. In our largest calculations, 7 pseudo-states were included in the two coupled channels and 5 pseudo-states were included in the $^3D_2$ channel.}\label{tab:pseudo-states}
\end{table}
To estimate to which degree excited states of the $^4$He target would influence our NCSM/RGM results for the $^6$Li g.s.\ energy, we performed a calculation that included g.s.\ and first excited $0^+$ state of $^4$He as well as deuteron g.s.\ and 7 $d^*$ pseudo states in the $^3S_1$-$^3D_1$ channel. The binding energy increased by 71 keV compared to the calculation with the same number of deuteron states but only the ground state of $^4$He. This is a non-negligible effect and, based on our previous study of the $n$-$^4$He system~\cite{NCSMRGM_PRC}, it is plausible that the addition of the next five or so lowest excited states of $^4$He would provide an extra $\sim500$ keV of binding. Unfortunately, such a calculation is currently out of reach.
\begin{figure}
\includegraphics*[width=0.9\columnwidth]{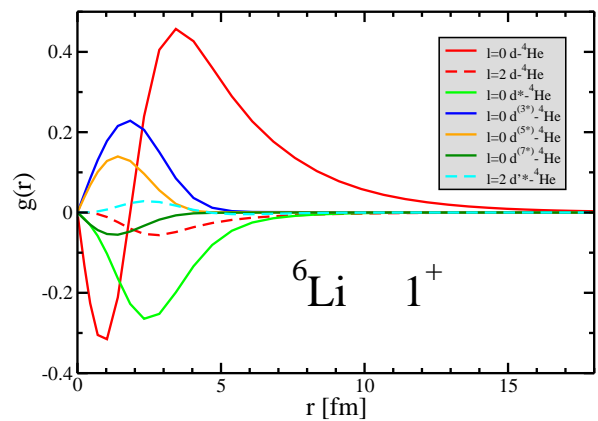}
\caption{(Color online) Ground-state wave function of $^6$Li as a function of the separation between deuteron and $^4$He clusters. Dominant $S$-wave (solid lines) and $D$-wave (dashed lines) components are shown. The symbol $d^*$ ($d^{(n*)}$) denotes the first ($n$-th) deuteron pseudo-state of the $^3S_1$-$^3D_1$ channel, while $d'^*$ denotes the lowest deuteron pseudo-state of the $^3D_3$-$^3G_3$ channel. Details of the calculation are described in the caption of Fig.~\protect\ref{phase_shift_d_alpha_best}.}\label{gs_wavefuction_d_alpha_best}
\end{figure}

Finally, we note that contrary to the NCSM calculations, the NCSM/RGM bound state has the proper asymptotic behavior of a Whittaker function with respect to the $d$+$\alpha$ threshold. In Fig.~\ref{gs_wavefuction_d_alpha_best}, we can see that the $S$-wave extends well beyond 10 fm.
This plot of the NCSM/RGM wave function can be compared with Fig. 6 of Ref.~\cite{cluster}. There the overlap functions [$g(r)/r$] of the $^6$Li ground state with $d$+$^4$He cluster states obtained within the standard NCSM 
vanish beyond about 8 fm.

\subsection{Scattering calculations}

By solving the NCSM/RGM coupled-channel equations (\ref{RGMeq}) for positive energies, we obtain the wave functions of the relative motion of the clusters and the scattering matrix for each considered $J^\pi T$ channel. The scattering matrix can then be used to calculate cross sections and other observables. 

\begin{figure}
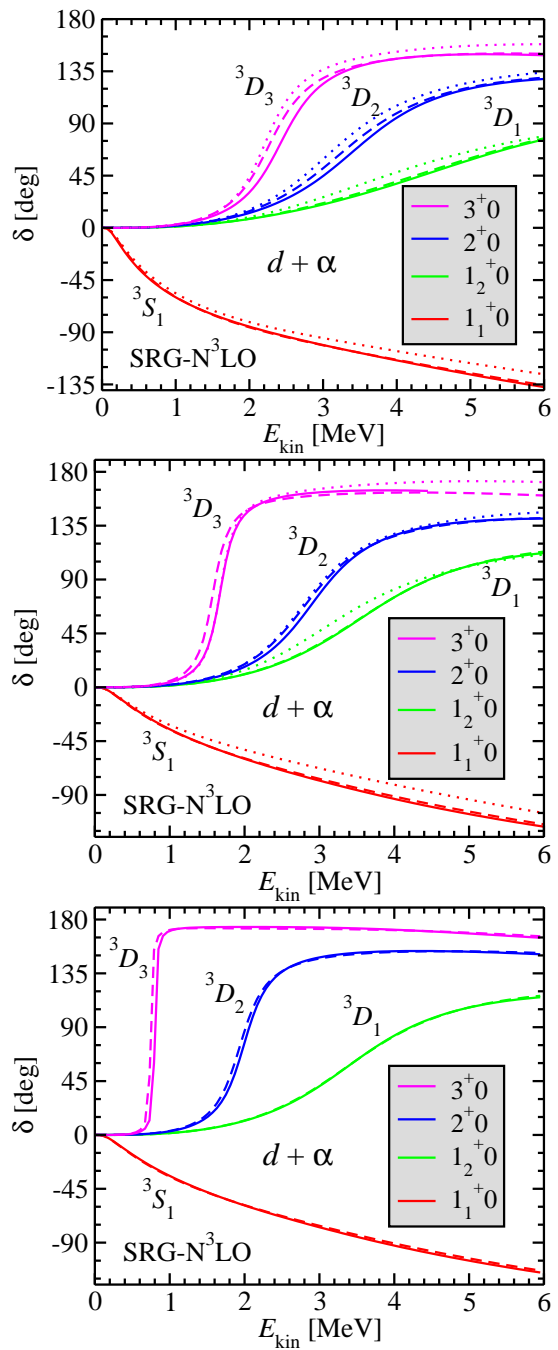

\begin{minipage}{8cm}
\includegraphics*[width=0.9\columnwidth]{d-He4_conv_srg-n3lo1p5_14_d0.eps}
\end{minipage}
\begin{minipage}{8cm}
\includegraphics*[width=0.9\columnwidth]{d-He4_conv_srg-n3lo1p5_14_d7.eps}
\end{minipage}
\begin{minipage}{8cm}
\includegraphics*[width=0.9\columnwidth]{d-He4_conv_srg-n3lo1p5_14_d7_3D37_3D27.eps}
\end{minipage}
\caption{(Color online) Calculated $d$-$^4$He $S$- and $D$-wave phase shifts. Solid, dashed and dotted lines correspond to the $N_{\rm max}=12, 10$ and $8$ basis sizes, respectively. Results in the top panel were obtained considering only the ground state of the deuteron projectile. In the middle panel, calculations incorporate 7 additional deuteron pseudo-states in the $^3S_1$-$^3D_1$ channel. In the bottom panel, up to 7 deuteron pseudo-states were included also in the $^3D_2$ and $^3D_3$-$^3G_3$ channels. The SRG-N$^3$LO $NN$ potential with $\Lambda=1.5$~ fm$^{-1}$ and the HO frequency of $\hbar\Omega=14$ MeV were used.}\label{phase_shift_d_alpha_conv}
\end{figure}

\begin{figure}
\includegraphics*[width=0.8\columnwidth]{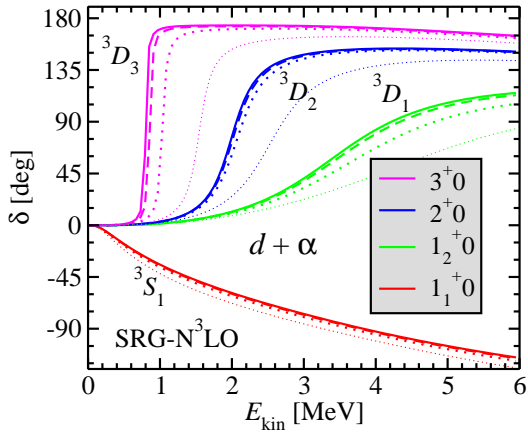}
\caption{(Color online) Calculated $d$-$^4$He $S$- and $D$-wave phase shifts. Dependence on the number of included deuteron pseudo-states in the  $^3S_1$-$^3D_1$, $^3D_2$ and $^3D_3$-$^3G_3$ channels: Solid, dashed, dotted, and thin dotted lines correspond to up to 7, 5, 3, and 1 pseudo-states in each channel, respectively. The SRG-N$^3$LO $NN$ potential with $\Lambda=1.5$~ fm$^{-1}$, the $N_{\rm max}=12$ basis size and the HO frequency of $\hbar\Omega=14$ MeV were used.}\label{phase_shift_d_alpha_dstar_conv}
\end{figure}

\begin{figure}
\includegraphics*[width=0.8\columnwidth]{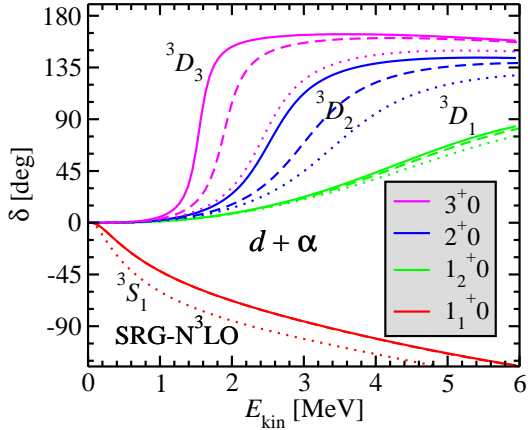}
\caption{(Color online) Calculated $d$-$^4$He $S$- and $D$-wave phase shifts. Influence of deuteron pseudo-states from different channels. Solid lines correspond to calculations with one deuteron pseudo-state in each of the $^3S_1$-$^3D_1$, $^3D_2$ and $^3D_3$-$^3G_3$ channels. Dashed lines identify results obtained with only one deuteron pseudo-state in the $^3S_1$-$^3D_1$ channel. Dotted-line curves are the solution of calculations including only the ground state of the deuteron. The SRG-N$^3$LO $NN$ potential with $\Lambda=1.5$~ fm$^{-1}$, the $N_{\rm max}=12$ basis size and the HO frequency of $\hbar\Omega=14$ MeV were used.}\label{phase_shift_d_alpha_channel_impact}
\end{figure}

\begin{figure}
\includegraphics*[width=0.8\columnwidth]{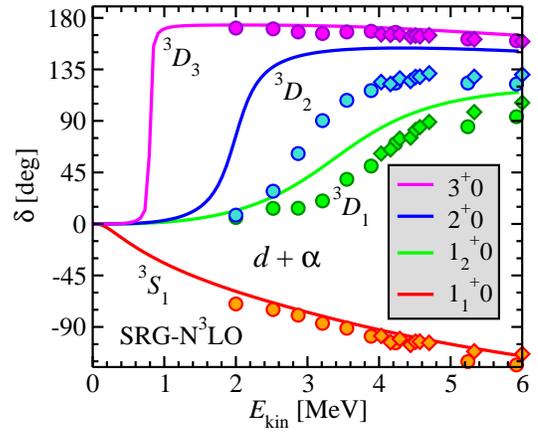}
\caption{(Color online) Calculated $d$-$^4$He $S$- and $D$-wave phase shifts compared to experimental data from Refs.~\protect\cite{Gruebler75} (circles) and~\protect\cite{Jenny83} (diamonds). Up to 7 deuteron pseudo-states were included in each of the $^3S_1$-$^3D_1$ and $^3D_3$-$^3G_3$ channels and 5 pseudo-states in the $^3D_2$ channel. The SRG-N$^3$LO $NN$ potential with $\Lambda=1.5$~ fm$^{-1}$, $N_{\rm max}=12$ basis size and HO frequency of $\hbar\Omega=14$ MeV were used.}\label{phase_shift_d_alpha_best}
\end{figure}

In Figs.~\ref{phase_shift_d_alpha_conv}-\ref{phase_shift_d_alpha_best} we present our calculated diagonal $S$- and $D$-wave phase shifts. First, we study the phase-shift convergence with respect to the size of the HO basis expansion for the cluster wave functions and localized parts of the integration kernels. In Fig.~\ref{phase_shift_d_alpha_conv}, we show phase shifts obtained  respectively in the $N_{\rm max}=12$ (solid line), $N_{\rm max}=10$ (dashed line), and $N_{\rm max}=8$ (dotted line) model spaces. The curves in the top panel 
include only the ground states of $d$ and $^4$He, while the middle and bottom panels show results including up to 7 deuteron pseudo-states in the $^3S_1$-$^3D_1$ and $^3D_2$ and $^3D_3$-$^3G_3$ channels. The $N_{\rm max}=10$ and $N_{\rm max}=12$ lines are on top of each other in the $J^\pi T=1^+ 0$ channels while some small change in the phase shifts is still visible in the $2^+ 0$ and the $3^+ 0$ channels. These differences become smaller in calculations with the pseudo-states. Overall, the convergence is satisfactory. At this stage an $N_{\rm max}=14$ calculation would be  computationally very challenging.
Figure~\ref{phase_shift_d_alpha_dstar_conv} demonstrates the phase shift convergence with respect to the number of pseudo-states included in the coupled-channel NCSM/RGM equations. It is clear that, similar to the bound-state calculation, for the $d^*$ channels considered here convergence is reached with 7 pseudo-states.
The relative contribution of  pseudo-states from the three $d^*$ channels considered here can be judged from Fig~\ref{phase_shift_d_alpha_channel_impact}. The $^3S_1$-$^3D_1$ pseudo-states affect all $S$ and $D$ waves. On the contrary, the $^3D_2$ and the $^3D_3$-$^3G_3$ pseudo-states have a considerable effect only on the $^3D_2$ and $^3D_3$ waves. Note that the solid lines in Fig.~\ref{phase_shift_d_alpha_channel_impact} correspond to the dotted lines in Fig.~\ref{phase_shift_d_alpha_dstar_conv}.

\begin{figure}
\includegraphics*[width=0.8\columnwidth]{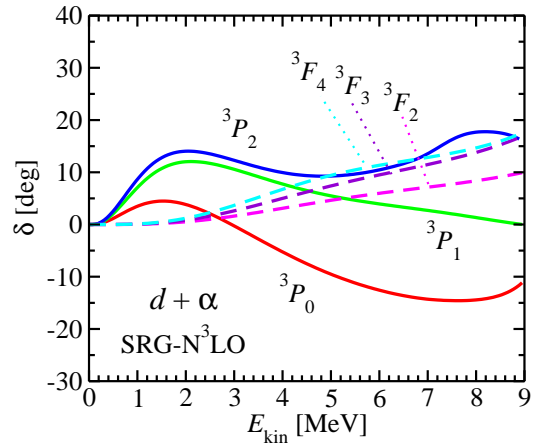}
\caption{(Color online) Calculated $d$-$^4$He $P$- and $F$-wave phase shifts. Details of the calculation are given in the caption of Fig.~\protect\ref{phase_shift_d_alpha_best}.}\label{phase_shift_d_alpha_P_F}
\end{figure}

Our calculated diagonal $S$- and $D$-wave phase shifts are compared to the phase shifts extracted from experimental data in Refs.~\cite{Gruebler75} and~\cite{Jenny83} in Fig.~\ref{phase_shift_d_alpha_best}. The calculation corresponds to the largest basis space ($N_{\rm amx}=12$) and the highest number of deuteron pseudo-states that we employed in this work. Our $S$-wave and $^3D_3$-wave results compare well with the experimental data. However, the $^3D_1$ and in particular the $^3D_2$ phase shifts overestimate the experimental ones. The position of our calculated $2^+ 0$ resonance is below the experimental one by almost 1 MeV. The splitting between the $D$-waves is underestimated. Clearly, the strength of the spin-orbit interaction in the calculation is smaller than it should be. This is most likely due to 
the neglect of the three-nucleon forces in our calculations, those induced by the SRG transformation and, 
more importantly, the initial chiral EFT three-nucleon interaction. 
Our calculated $P$- and $F$-wave phase shifts are presented in Fig.~\ref{phase_shift_d_alpha_P_F}. While the $F$-waves monotonically increase, the $P$-waves exhibit more structure and, in particular, the $^3P_0$ changes sign and becomes negative beyond the center-of-mass energy $E_{\rm kin}> 2$~MeV.

\begin{figure}
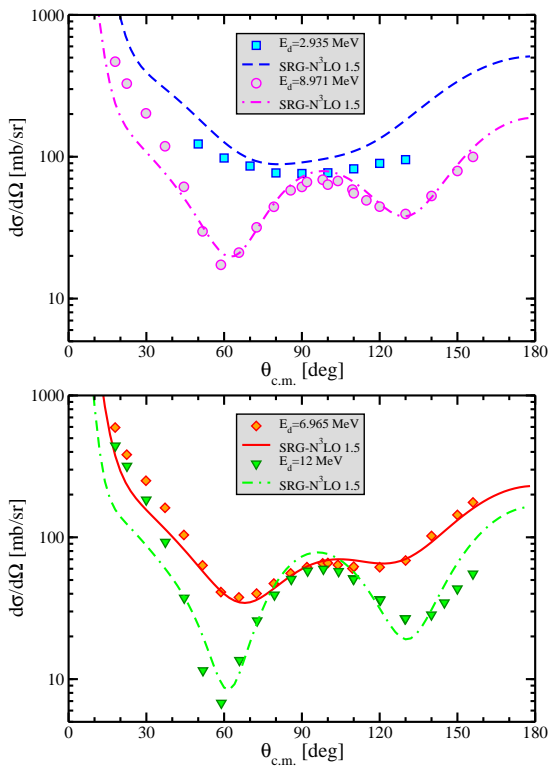

\begin{minipage}{8cm}
\includegraphics*[width=0.9\columnwidth]{dsigma_dOmega_dHe4_srg-n3lo1p5_14_13_d7_3D37_3D27_1p95MeV_5p95MeV_fig.eps}
\end{minipage}
\begin{minipage}{8cm}
\includegraphics*[width=0.9\columnwidth]{dsigma_dOmega_dHe4_srg-n3lo1p5_14_13_d7_3D37_3D27_4p6MeV_8MeV_fig.eps}
\end{minipage}
\caption{(Color online) The $d$-$^4$He differential cross section at the deuteron laboratory energies of 2.935, 6.965, 8.971 and 12 MeV. The experimental data (symbols) are from Ref.~\protect\cite{Senhouse64} and~\cite{Jett77}. The calculations (lines) are as described in Fig.~\protect\ref{phase_shift_d_alpha_best}. Partial waves up to $J{=}6$ were included.}\label{ds_dOm_d_alpha_best}
\end{figure}

\begin{figure}
\includegraphics*[width=0.9\columnwidth]{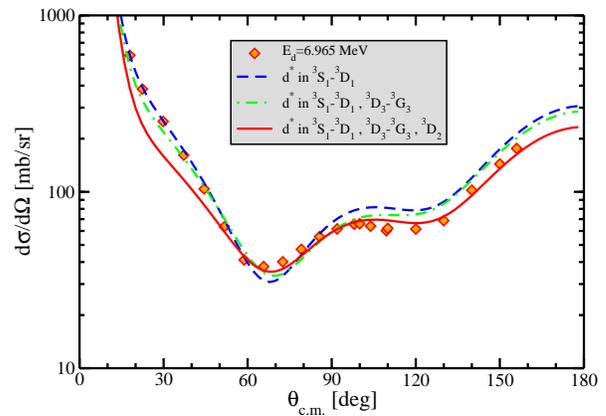}
\caption{(Color online) Influence of deuteron pseudo-states from different channels on the $d$-$^4$He differential cross section at the deuteron laboratory energy of 6.965 MeV. The experimental data (symbols) are from Ref.~\protect\cite{Senhouse64}.  The solid line corresponds to the calculation described in Fig.~\protect\ref{phase_shift_d_alpha_best}. The dashed-dotted line indicates the results obtained with 7 deuteron pseudo-states in  the $^3S_1$-$^3D_1$ and $^3D_3$-$^3G_3$ channels. The dashed-line curve is the solution of the calculation with 7 deuteron pseudo-states only in the  $^3S_1$-$^3D_1$ channel. Partial waves up to $J{=}6$ were included.}\label{ds_dOm_d_alpha_channelcontr}
\end{figure}

In Fig.~\ref{ds_dOm_d_alpha_best}, we compare our calculated differential cross section to the experimental data of Refs.~\cite{Senhouse64} and~\cite{Jett77} for four deuteron laboratory energies in the range $E_d\approx 3-12$~MeV. Our calculation over-predicts the measured cross section at $E_d=2.94$ MeV, most likely a consequence of the incorrect position of the calculated $2^+ 0$ resonance, see Fig~\ref{phase_shift_d_alpha_best}. However, for the intermediate energies, $E_d=6.97$ and 8.97~MeV, the agreement with the measured data is reasonable. At $E_d=12$~MeV the differences become larger. We also note that our calculated cross section underestimates the data in the range of $\theta_{\rm c.m.}\approx 20-45$~deg. It should be kept in mind that in our calculations the deuteron breakup is accounted for only by using the pseudo-states rather than as a three-body final state. To shed light on the influence of the pseudo-states on the cross sections, we show in Fig.~\ref{ds_dOm_d_alpha_channelcontr} results for $E_d=6.97$~MeV obtained with pseudo-states in different channels. With the pseudo-states only in the $^3S_1{-}^3D_1$ channel, the cross section is not well described beyond 50 deg. The inclusion of the pseudo-states in the $^3D_3$-$^3G_3$ channel improves the agreement with the data somewhat. By adding the pseudo-states in the $^3D_2$ channel, the agreement with the data beyond 50 deg is quite reasonable but at the forward angles, from 20 deg to 45 deg, the agreement with the data is spoiled. The $NN$ interaction that we employed is not the optimal one as explained earlier: A low value of $\Lambda=1.5$~fm$^{-1}$ was selected to facilitate a fast convergence and a straightforward comparison to the standard NCSM calculation. As seen in Fig.~\ref{phase_shift_d_alpha_best}, this potential overestimates the $2^+ 0$ $D$-wave phase shifts compared to the data. The pseudo-states from the $^3D_2$ channel enhance this overestimation as visible in Fig.~\ref{phase_shift_d_alpha_channel_impact}. This is the likely cause of the worsening of the cross-section agreement with the data at forward angles when the $^3D_2$ pseudo-states are added to the NCSM/RGM basis.

\begin{figure}
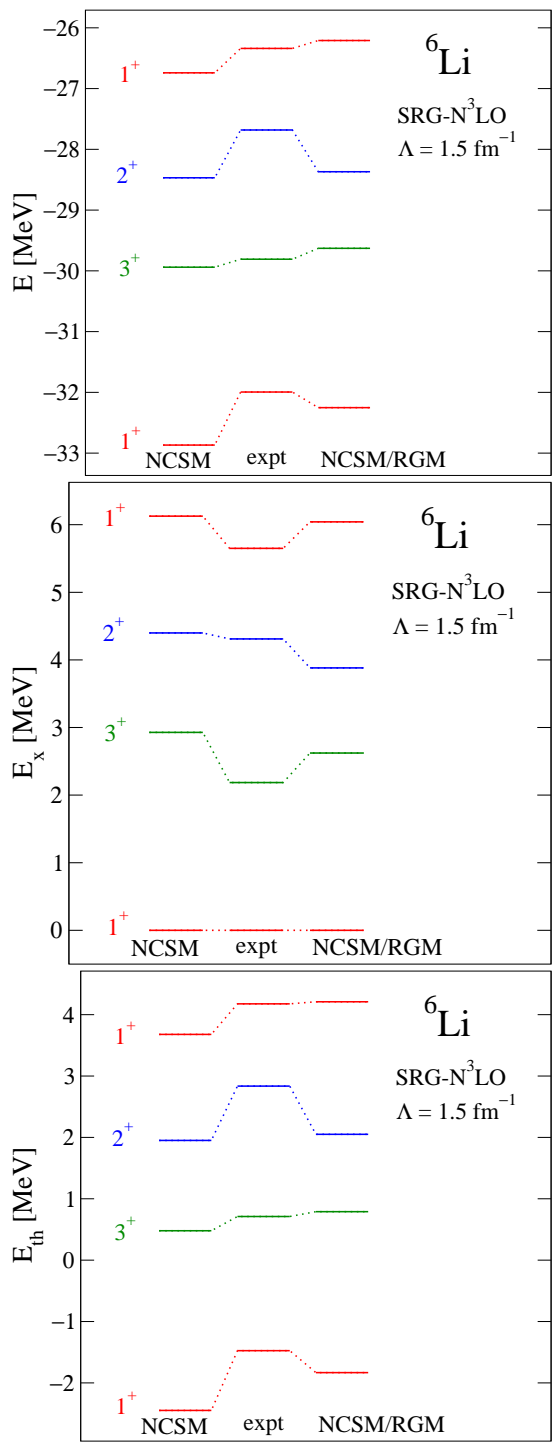

\begin{minipage}{8cm}
\includegraphics*[width=0.9\columnwidth]{Li6_srg-n3lo_1p5_14_NCSM_vs_NCSMRGM.eps}
\end{minipage}
\begin{minipage}{8cm}
\includegraphics*[width=0.9\columnwidth]{Li6_srg-n3lo_1p5_14_NCSM_vs_NCSMRGM_excited.eps}
\end{minipage}
\begin{minipage}{8cm}
\includegraphics*[width=0.9\columnwidth]{Li6_srg-n3lo_1p5_14_NCSM_vs_NCSMRGM_thres.eps}
\end{minipage}
\caption{(Color online) Energies of $^6$Li states calculated within NCSM and NCSM/RGM are compared to each other and to experimental data. Absolute energies (top panel), excitations energies (middle panel) and energies with respect to calculated and experimental $d$-$^4$He threshold (bottom panel) are presented. Details of the calculations are given in Table~\protect\ref{tab:Egs_2-6}.}\label{NCSMvsNCSMRGM}
\end{figure}

The g.s.\ and the $T=0$ resonance-state energies obtained within the NCSM and the NCSM/RGM are compared to each other and to experimental values in Fig.~\ref{NCSMvsNCSMRGM}. The present NCSM/RGM resonance energies correspond to the energies where the diagonal phase shifts cross 90 degrees. The calculation is as described in Table~\ref{tab:Egs_2-6} and Fig.~\ref{phase_shift_d_alpha_best}. We plot the absolute values of the energies as well as the excitation energies and the energies relative to the calculated and experimental thresholds. Overall, the NCSM calculation produces more binding by about 600 keV as already discussed in the previous subsection. The NCSM/RGM generates excitation energies for the resonances systematically lower than the corresponding NCSM results. There is in particular a significant shift for the $2^+ 0$ state. At the same time, as it can be seen from the bottom panel of Fig.~\ref{Li6_NCSM}, the excitation energies of the $2^+ 0$ and the $1^+ 0$ states show a slower convergence rate with respect to the size of the HO basis expansion. This is a consequence of the inadequacy of the HO basis for the description of broader resonances. In this regard, the NCSM/RGM calculation is clearly superior.

\section{Conclusions}
\label{conclusions}

In this paper, we extended the {\it ab initio} NCSM/RGM approach to projectile-target binary-cluster states where the projectile is a deuteron. 
We gave details on the new aspects of the formalism and presented algebraic expressions for the integration kernels for the specific case  in which the target wave functions are expanded in the SD HO basis. Among the new features, the dependence of the Hamiltonian kernel upon the three-body density of the target makes calculations technically challenging, due to the rapidly increasing number of matrix elements with the size of the multi-major shell basis. 

To test our formalism, we performed calculations for the bound and scattering states of the $d$-$\alpha$ system. In this case the three-body density calculation can be performed in a straightforward way using a closure relation. As the deuteron is weakly bound, its polarization and breakup cannot be neglected. A proper treatment would require the inclusion in the NCSM/RGM formalism of three-body final states: $n$-$p$-$\alpha$. Although extensions of the approach in this direction are possible, this is quite challenging and has not been explored yet. In this first application, we approximated the three-body continuum by using deuteron pseudo-states. We compared our $d$-$^4$He results to experimental data as well as to a standard NCSM calculation for $^6$Li using the same Hamiltonian. To facilitate benchmarking with the standard NCSM, we employed a soft SRG-N$^3$LO $NN$ potential with a low evolution parameter $\Lambda=1.5$~fm$^{-1}$. In this way, we were able to reach convergence at $N_{\rm max}\approx 12$. The differences between the NCSM/RGM and the NCSM results are then due to omitted correlations rather than to the adopted HO basis size. Interestingly, the NCSM calculation produced a $^6$Li g.s.\ energy lower by about 2\%.  
This means that internal excitations of the $^4$He target, neglected in our NCSM/RGM calculation, play some role and/or that the pseudo-state approximation of the three-body continuum is not completely adequate. On the other hand, the NCSM/RGM calculation generates lower excitation energies for the broader resonances that present a slower convergence rate with respect to the HO basis expansion. 

Overall, the NCSM/RGM calculation is superior to the standard NCSM because it generates wave functions with proper boundary conditions for the bound state and, further, describes resonances and scattering states. However, to include all relevant excitations is a challenge. Therefore, clearly the way forward is a unification of the two approaches. This can be accomplished by coupling the present NCSM/RGM basis, consisting of binary-cluster channels with just a few lowest excited states of projectile and target, with the NCSM eigenstates of the composite system as outlined in Ref.~\cite{NCSM_review}. Work on this unified approach is under way.

Our immediate plans include the application of the NCSM/RGM formalism
to the $^3$H($d$,$n$)$^4$He and $^3$He($d$,$p$)$^4$He fusion reactions. This requires working in a NCSM/RGM model space including both $n$-$^4$He ($p$-$^4$He) and $d$-$^3$H ($d$-$^3$He) channel states, that is a deuteron-nucleon $(d,N)$ transfer formalism which combines  the deuteron-nucleus (presented here) and nucleon-nucleus (presented in Ref.~\cite{NCSMRGM_PRC}) formalisms as well as the integration kernels resulting from the coupling between the $(A{-}1,1)$ and $(A{-}2,2)$ mass partitions, which will be the subject of a forthcoming publication.
Our preliminary $^3$He($d$,$p$)$^4$He S-factor results were discussed in Ref.~\cite{NCSM_RGM_INPC10}.

The use of SRG-evolved $NN$ interaction facilitates the convergence of the NCSM/RGM calculations with respect to the HO basis expansion. On the other hand, due to the softness of these interactions, radii of heavier nuclei become underestimated. Therefore, it is essential to further develop the NCSM/RGM formalism in order to handle three-nucleon interactions, both genuine and SRG-evolution induced, in bound-state and scattering calculations.

To apply the present deuteron-nucleus formalism to heavier target nuclei, i.e. heavy $p$-shell nuclei and beyond, it becomes necessary to utilize the recently developed importance-truncated NCSM~\cite{IT-NCSM,Roth09}. This gives us the ability to use large $N_{\rm max}$ model spaces, that in the NCSM/RGM approach are of vital importance not just for the convergence of the target and projectile eigenstates but also for the convergence of the localized parts of the integration kernels~\cite{NCSMRGM_IT}.

Finally, our future plans also include a further generalization of the formalism to projectile-target binary-cluster states with three-nucleon ($^3$H, $^3$He) and four-nucleon ($^4$He) projectiles. Calculations of the integrations kernels for the three-nucleon projectile case are under way.

\acknowledgments
Computing support for this work came from the Lawrence Livermore National Laboratory (LLNL) Institutional Computing Grand Challenge program.
Prepared in part by LLNL under Contract DE-AC52-07NA27344.
Support from the LLNL LDRD grant PLS-09-ERD-020 is acknowledged.

\begin{widetext}
\appendix
\section{Hamiltonian kernel matrix elements}
\label{appA}
Here we present matrix elements of the potential kernel (\ref{V-kernel}) in the basis states (\ref{SD-basis-ab}). For the first term on the rhs of Eq.~(\ref{antisym_HAm2}), we obtain
\begin{align}
 &\leftsub{\rm SD}{\left\langle\Phi^{J^\pi T}_{\kappa_{ab}^\prime}\left| V_{A-2,A-1}(1-\hat{P}_{A-2,A-1}) \right|\Phi^{J^\pi T}_{\kappa_{ab}}\right\rangle}_{\rm SD} \nonumber\\[2mm]
 & \qquad =  
\delta_{a,a^\prime}\frac{1}{A-2}{\sum_{c\,c^\prime} \sum_{J_{bc} T_{bc}} \sum_{K\,\tau}}
\left\{ \begin{array}{@{\!~}c@{\!~}c@{\!~}c@{\!~}} 
I_1 & K & I_1^\prime \\[2mm] 
I^\prime & J & I 
\end{array}\right\}
\left\{ \begin{array}{@{\!~}c@{\!~}c@{\!~}c@{\!~}} 
j_b & j_b^\prime & K \\[2mm] 
I^\prime & I & j_a 
\end{array}\right\}
\left\{ \begin{array}{@{\!~}c@{\!~}c@{\!~}c@{\!~}} 
j_b^\prime & j_c^\prime & J_{bc} \\[2mm] 
j_c & j_b & K 
\end{array}\right\}
\left\{ \begin{array}{@{\!~}c@{\!~}c@{\!~}c@{\!~}} 
T_1 & \tau & T_1^\prime \\[2mm] 
T_2^\prime & T & T_2 
\end{array}\right\}
\left\{ \begin{array}{@{\!~}c@{\!~}c@{\!~}c@{\!~}} 
\frac{1}{2} & \frac{1}{2} & \tau \\[2mm] 
T_2^\prime & T_2 & \frac{1}{2} 
\end{array}\right\}
\left\{ \begin{array}{@{\!~}c@{\!~}c@{\!~}c@{\!~}} 
\frac{1}{2} & \frac{1}{2} & T_{bc} \\[2mm] 
\frac{1}{2} & \frac{1}{2} & \tau
\end{array}\right\}
\nonumber \\[2mm]
&\qquad \phantom{=} \times 
(-1)^{J_{bc}+j_a+j_c^\prime+K-I_1-J} \; (-1)^{T_{bc}+\tau+1-T_1-T}\, \hat{I}\, \hat{I}^\prime \, \hat{K}
\, \hat{T}_2\, \hat{T}_2^\prime \, \hat{\tau}\,  \hat{J}_{bc}^2\,  \hat{T}_{bc}^2
\nonumber \\[2mm]
&\qquad \phantom{=}\times 
\leftsub{\rm SD}{ \left\langle A{-}2\, \alpha_1^\prime I_1^\prime T_1^\prime \left | \left | \left | 
\left(a^\dagger_{c^\prime}\tilde{a}^{\phantom l}_{c}\right)^{(K\tau)} \right | \right | \right | A{-}2\, \alpha_1 I_1 T_1\right \rangle}_{\rm SD}
\,\sqrt{1+\delta_{b^\prime,c^\prime}}\, \sqrt{1+\delta_{b,c}}\;\;
\left\langle b^\prime c^\prime J_{bc} T_{bc} \left | V \right | b \,c \,J_{bc} T_{bc} \right\rangle
\,,
\label{V-kernel_1}
\end{align}
where we abbreviate $a\equiv n_a l_a j_a \textstyle{\frac{1}{2}}$ etc.  We note that the matrix elements of the interaction $V$ in the antisymmetrized and normalized two-body basis are evaluated using just the first term of Eq.~(\ref{pot}), i.e. 
$V_{ij}=V_N(ij)+\frac{e^2(1+\tau^z_i)(1+\tau^z_j)}{4|\vec r_i-\vec r_j|}$ (with $V_N$ the nuclear part)
as the average Coulomb interaction is taken care of with the help of Eq.~(43) in Ref~\cite{NCSMRGM_PRC}. 
For the second term on the rhs of Eq.~(\ref{antisym_HAm2}) we derive
\begin{align}
&\leftsub{\rm SD}{\left \langle\Phi^{J^\pi T}_{\kappa_{ab}^\prime} \left | V_{A-2,A}\hat{P}_{A-2,A-1} \right| \Phi^{J^\pi T}_{\kappa_{ab}}\right\rangle}_{\rm SD}  
\nonumber\\[2mm]
& \qquad =
\frac{1}{2}\frac{1}{A-2}{\sum_{c^\prime K \tau}}
\left\{ \begin{array}{@{\!~}c@{\!~}c@{\!~}c@{\!~}} 
I_1 & K & I_1^\prime \\[2mm] 
I^\prime & J & I 
\end{array}\right\}
\left\{ \begin{array}{@{\!~}c@{\!~}c@{\!~}c@{\!~}} 
j_c^\prime & j_b^\prime & K \\[2mm] 
I^\prime & I & j_a^\prime 
\end{array}\right\}
\left\{ \begin{array}{@{\!~}c@{\!~}c@{\!~}c@{\!~}} 
T_1 & \tau & T_1^\prime \\[2mm] 
T_2^\prime & T & T_2 
\end{array}\right\}
\left\{ \begin{array}{@{\!~}c@{\!~}c@{\!~}c@{\!~}} 
\frac{1}{2} & \frac{1}{2} & \tau \\[2mm] 
T_2^\prime & T_2 & \frac{1}{2} 
\end{array}\right\}
(-1)^{j_b^\prime+j_a^\prime+K-I_1-J} \; (-1)^{\tau+1-T_1-T}
\, \hat{I}\, \hat{I}^\prime \, \hat{K}
\, \hat{T}_2 \, \hat{T}_2^\prime\,  \hat{\tau}
\nonumber \\[2mm]
&\qquad \phantom{=}\times 
\leftsub{\rm SD}{\left\langle A{-}2\, \alpha_1^\prime I_1^\prime T_1^\prime \left | \left | \left | 
\left(a^\dagger_{c^\prime}\tilde{a}^{\phantom l}_{b^\prime}\right)^{(K\tau)} \right | \right | \right |A{-}2\, \alpha_1 I_1 T_1\right\rangle}_{\rm SD}
\;\sqrt{1+\delta_{a^\prime,c^\prime}}\, \sqrt{1+\delta_{a,b}}\;\;
\left \langle a^\prime c^\prime I T_2 \left | V \right | a \, b\,  I T_2 \right \rangle
\,.
\label{V-kernel_2}
\end{align}
%
For the third term we get
%
\begin{align}
&\leftsub{\rm SD}{\left \langle\Phi^{J^\pi T}_{\kappa_{ab}^\prime} \left | V_{A-3,A}(1-\hat{P}_{A-3,A}) \hat{P}_{A-2,A-1} 
\right |\Phi^{J^\pi T}_{\kappa_{ab}}\right\rangle}_{\rm SD}
\nonumber\\
&\qquad =\frac{1}{(A-2)(A-3)}{\sum_{d\,d^\prime} \sum_{J_{ad}T_{ad}}\sum_{K_1 K_2 K} \sum_{\tau_1 \tau_2 \tau}}
\left\{ \begin{array}{@{\!~}c@{\!~}c@{\!~}c@{\!~}} 
I_1 & K & I_1^\prime \\[2mm] 
I^\prime & J & I 
\end{array}\right\}
\left\{ \begin{array}{@{\!~}c@{\!~}c@{\!~}c@{\!~}} 
T_1 & \tau & T_1^\prime \\[2mm] 
T_2^\prime & T & T_2 
\end{array}\right\}
\left\{ \begin{array}{@{\!~}c@{\!~}c@{\!~}c@{\!~}c@{\!~}} 
K_1 & K_2 & j_d & j_a \\[2mm] 
K  & j_b^\prime & J_{ad} & j_b \\[2mm] 
I   & I^\prime & j_a^\prime & j_d^\prime \\[2mm] 
\end{array}\right\}
\left\{ \begin{array}{@{\!~}c@{\!~}c@{\!~}c@{\!~}c@{\!~}} 
\tau_1 & \tau_2 & \frac{1}{2} & \frac{1}{2} \\[2mm] 
\tau  & \frac{1}{2} & T_{ad} & \frac{1}{2} \\[2mm] 
T_2  & T_2^\prime & \frac{1}{2} & \frac{1}{2} \\[2mm] 
\end{array}\right\}
\nonumber \\[2mm]
&\qquad\phantom{=}\times
(-1)^{I_1+J+j_b-j_d+J_{ad}}\; (-1)^{T_1+T+T_{ad}} \, \hat{I}\, \hat{I}^\prime \, \hat{K} \, \hat{K}_1\,  \hat{K}_2 
\, \hat{T}_2\, \hat{T}_2^\prime\,  \hat{\tau}\,  \hat{\tau}_1\, \hat{\tau}_2 \, \hat{J}_{ad}^2\, \hat{T}_{ad}^2
\nonumber \\[2mm]
&\qquad\phantom{=}\times
 \leftsub{\rm SD}{\left \langle A{-}2\, \alpha_1^\prime I_1^\prime T_1^\prime \left | \left | \left | 
\left((a^\dagger_b a^\dagger_{d^\prime})^{(K_1 \tau_1)}(\tilde{a}_{d}\tilde{a}_{b^\prime})^{(K_2 \tau_2)}\right)^{(K\tau)}\right | \right | \right |  A{-}2\, \alpha_1 I_1 T_1\right\rangle}_{\rm SD}
\nonumber \\[2mm]
&\qquad\phantom{=}\times
\sqrt{1+\delta_{a^\prime,d^\prime}}\, \sqrt{1+\delta_{a,d}}\;\;
\left\langle a^\prime d^\prime J_{ad} T_{ad} \left | V \right | a\, d\, J_{ad} T_{ad} \right\rangle 
\,,
\label{V-kernel_3}
\end{align}
with the $12{-}j$ symbols of the first kind~\cite{Varshalovich} in the notation and the definition given in Appendix A of Ref.~\cite{cluster}.

The matrix element of the fourth term on the rhs of Eq.~(\ref{antisym_HAm2}) is obtained in the form
\begin{align}
& \leftsub{\rm SD}{\left\langle\Phi^{J^\pi T}_{\kappa_{ab}^\prime}\left| V_{A-3,A-1}\hat{P}_{A-2,A-1} \right|\Phi^{J^\pi T}_{\kappa_{ab}}\right\rangle}_{\rm SD}
\nonumber\\
 &\qquad =
\delta_{a,a^\prime}\frac{1}{2}\frac{1}{(A-2)(A-3)}{\sum_{c\,d\,d^\prime} \sum_{J_{bd^\prime} T_{bd^\prime}} \sum_{J_{cd} T_{cd}} \sum_{K\,\tau}}
\left\{ \begin{array}{@{\!~}c@{\!~}c@{\!~}c@{\!~}} 
I_1 & K & I_1^\prime \\[2mm] 
I^\prime & J & I 
\end{array}\right\}
\left\{ \begin{array}{@{\!~}c@{\!~}c@{\!~}c@{\!~}} 
j_b & j_b^\prime & K \\[2mm] 
I^\prime & I & j_a 
\end{array}\right\}
\left\{ \begin{array}{@{\!~}c@{\!~}c@{\!~}c@{\!~}} 
j_b^\prime & j_d^\prime & J_{cd} \\[2mm] 
J_{bd^\prime} & K & j_b 
\end{array}\right\}
\nonumber \\
& \qquad\phantom{=}\times 
\left\{ \begin{array}{@{\!~}c@{\!~}c@{\!~}c@{\!~}} 
T_1 & \tau & T_1^\prime \\[2mm] 
T_2^\prime & T & T_2 
\end{array}\right\}
\left\{ \begin{array}{@{\!~}c@{\!~}c@{\!~}c@{\!~}} 
\frac{1}{2} & \frac{1}{2} & \tau \\[2mm] 
T_2^\prime & T_2 & \frac{1}{2} 
\end{array}\right\}
\left\{ \begin{array}{@{\!~}c@{\!~}c@{\!~}c@{\!~}} 
\frac{1}{2} & \frac{1}{2} & T_{cd} \\[2mm] 
T_{bd^\prime} & \tau & \frac{1}{2}
\end{array}\right\}
\nonumber \\[2mm]
& \qquad\phantom{=}\times
(-1)^{j_a+j_b+j_c+j_d+j_b^\prime+j_d^\prime+I_1+J} \; (-1)^{1+T_1+T}\,\hat{I}\,\hat{I}^\prime \,\hat{K}
\,\hat{T}_2\,\hat{T}_2^\prime \,\hat{\tau} \,\hat{J}_{cd} \,\hat{J}_{bd^\prime} \,\hat{T}_{cd} \,\hat{T}_{bd^\prime}
\nonumber \\[2mm]
& \qquad\phantom{=}\times
 \leftsub{\rm SD}{\left\langle A{-}2\, \alpha_1^\prime I_1^\prime T_1^\prime \left|\left|\left| 
\left((a^\dagger_b a^\dagger_{d^\prime})^{(J_{bd^\prime}T_{bd^\prime})}(\tilde{a}_{d}\tilde{a}_{c})^{(J_{cd} T_{cd})}\right)^{(K\tau)}\right|\right|\right|A{-}2\, \alpha_1 I_1 T_1\right\rangle}_{\rm SD}
\nonumber \\[2mm]
& \qquad\phantom{=}\times
\sqrt{1+\delta_{b^\prime,d^\prime}} \, \sqrt{1+\delta_{c,d}}\;\;
\left\langle b^\prime d^\prime J_{cd} T_{cd} \left| V \right| c\, d\, J_{cd} T_{cd} \right\rangle 
\,.
\label{V-kernel_4}
\end{align}

Finally, for the last term on the rhs of Eq.~(\ref{antisym_HAm2}) we find
%
\begin{align}
& \leftsub{\rm SD}{\left\langle\Phi^{J^\pi T}_{\kappa_{ab}^\prime}\left|V_{A,A-4} \hat{P}_{A-2,A-1}\hat{P}_{A-3,A} \right|\Phi^{J^\pi T}_{\kappa_{ab}}\right\rangle}_{\rm  SD}
\nonumber\\
 & \qquad =
\frac{1}{2}\frac{1}{(A-2)(A-3)(A-4)}{\sum_{d\,e\,e^\prime} \sum_{J_{de} T_{de}} \sum_{K_1 \tau_1} \sum_{K\, \tau}}
\left\{ \begin{array}{@{\!~}c@{\!~}c@{\!~}c@{\!~}} 
I_1 & K & I_1^\prime \\[2mm] 
I^\prime & J & I 
\end{array}\right\}
\left\{ \begin{array}{@{\!~}c@{\!~}c@{\!~}c@{\!~}} 
T_1 & \tau & T_1^\prime \\[2mm] 
T_2^\prime & T & T_2 
\end{array}\right\}
\left\{ \begin{array}{@{\!~}c@{\!~}c@{\!~}c@{\!~}} 
J_{de} & j_e^\prime & j_a^\prime \\[2mm] 
j_{b^\prime} & I^\prime & K_1 
\end{array}\right\}
\left\{ \begin{array}{@{\!~}c@{\!~}c@{\!~}c@{\!~}} 
T_{de} & \frac{1}{2} & \frac{1}{2} \\[2mm] 
 \frac{1}{2} & T_2^\prime & \tau_1 
\end{array}\right\}
\nonumber \\[2mm]
&\qquad\phantom{=} \times
(-1)^{I-I_1-J+K_1+j_e+j_d+j_b^\prime+j_a^\prime} \; (-1)^{T_2-T_1-T+\tau_1}
\,\hat{K}\, \hat{\tau}\, \hat{K}_1\, \hat{J}_{de}\, \hat{\tau}_1\, \hat{T}_{de}
\nonumber \\[2mm]
&\qquad\phantom{=} \times
 \leftsub{\rm SD}{\left\langle A{-}2\, \alpha_1^\prime I_1^\prime T_1^\prime \left|\left|\left| 
\left((a^\dagger_a a^\dagger_b)^{(I T_2)}\left((a^\dagger_{e^\prime}\tilde{a}^{\phantom l}_{b^\prime})^{(K_1 \tau_1)} (\tilde{a}_{e}\tilde{a}_{d})^{(J_{de} T_{de})}\right)^{(I^\prime T_2^\prime)}\right)^{(K\tau)}\right|\right|\right|A{-}2\, \alpha_1 I_1 T_1\right\rangle}_{\rm SD}
\nonumber \\[2mm]
& \qquad\phantom{=}\times
\sqrt{1+\delta_{a^\prime,e^\prime}} \, \sqrt{1+\delta_{d,e}}\;\;
\left\langle a^\prime e^\prime J_{de} T_{de} \left| V \right| d \,e\, J_{de} T_{de} \right\rangle 
\,.
\label{V-kernel_5}
\end{align}

\end{widetext}


\begin{thebibliography}{10}

\bibitem{Witala01} H. Witala, W. Gl\"ockle, J. Golak, A. Nogga, H. Kamada, R. Skibinski, 
     and J. Kuros-Zolnierczuk, Phys. Rev. C {\bf 63}, 024007 (2001).

\bibitem{Lazauskas05} R. Lazauskas and J. Carbonell, Phys. Rev. C {\bf 70}, 044002 (2004).

\bibitem{Pisa} A. Kievsky, S. Rosati, M. Viviani, L. E. Marcucci and L. Girlanda, 
               J. Phys. G {\bf 35}, 063101 (2008).

\bibitem{Deltuva} A.\ Deltuva and A.\ C.\ Fonseca, 
                      Phys.\ Rev.\ C {\bf 75}, 014005 (2007);  
                      Phys.\ Rev.\ Lett. {\bf 98}, 162502 (2007).

\bibitem{GFMC_nHe4} K.\ M.\ Nollett, S. C. Pieper, R. B. Wiringa, 
                    J. Carlson and G. M. Hale, 
                    Phys.\ Rev.\ Lett.\ {\bf 99}, 022502 (2007). 

\bibitem{RGM} K.\ Wildermuth and Y.\ C.\ Tang, {\it A unified theory of the nucleus},
              (Vieweg, Braunschweig, 1977). 

\bibitem{RGM1} Y.\ C.\ Tang, M. LeMere and D. R. Thompson,
              Phys.\ Rep.\ {\bf 47}, 167 (1978).
             
\bibitem{RGM2} T. Fliessbach and H. Walliser, Nucl. Phys. {\bf A377}, 84 (1982).

\bibitem{RGM3} K.\ Langanke and H.\ Friedrich, {\it Advances in Nuclear Physics}, 
              edited by J. W. Negele and E. Vogt (Plenum, New York, 1986).

\bibitem{Lovas98} R. G. Lovas, R. J. Liotta, A. Insolia, K. Varga and D. S. Delion,
              Phys. Rep. {\bf 294}, 265 (1998).

\bibitem{Hofmann08} H. M. Hofmann and G. M. Hale, Phys. Rev. C {\bf 77}, 044002 (2008).

\bibitem{NCSMC12} P.\ Navr\'atil, J.\ P.\ Vary, and B.\ R.\ Barrett,
                   Phys.\ Rev.\ Lett. {\bf 84}, 5728 (2000);
                   Phys.\ Rev.\ C {\bf 62}, 054311 (2000).

\bibitem{NCSMRGM} S.\ Quaglioni and P.\ Navr{\'a}til,  Phys. Rev. Lett. {\bf 101}, 092501 (2008). 

\bibitem{NCSMRGM_PRC} S.\ Quaglioni and P.\ Navr{\'a}til,  Phys. Rev. C  {\bf 79}, 044606 (2009).

\bibitem{NCSMRGM_IT} P.\ Navr{\'a}til, R.\ Roth and S.\ Quaglioni, Phys. Rev. C  {\bf 82}, 034609 (2010).

\bibitem{3bbound1} P.\ Descouvemont, C.\ Daniel, and D.\ Baye, Phys.\ Rev.\  C {\bf 67}, 044309 (2003).

\bibitem{3bcont2} D.\ Baye, P.\ Capel, P.\ Descouvemont, and Y.\ Suzuki, Phys. Rev. C {\bf 79}, 024607 (2009).

\bibitem{Thompson73} D. R. Thompson and Y. C. Tang, Phys. Rev. C {\bf 8}, 1649 (1973).

\bibitem{Kanada82} H. Kanada, T. Kaneko and Y. C. Tang, Nucl. Phys. A {\bf 389}, 285 (1982).

\bibitem{Kanada85} H. Kanada, T. Kaneko, S. Saito and Y. C. Tang, Nucl. Phys. A {\bf 444}, 209 (1985).

\bibitem{SRG} S. K. Bogner, R. J. Furnstahl and R. J. Perry, Phys. Rev. C {\bf 75}, 061001 (2007).

\bibitem{Roth_SRG} R. Roth, S. Reinhardt and H. Hergert, Phys. Rev. C {\bf 77}, 064003 (2008). 

\bibitem{N3LO} D.\ R.\ Entem and R.\ Machleidt, Phys.\ Rev.\ C {\bf 68}, 041001(R) (2003).

\bibitem{R-matrix} M.\ Hesse, J.-M. Sparenberg, F. Van Raemdonck, and D. Baye,
                                 Nucl.\ Phys.\ {\bf A640}, 37 (1998); 
                                 M.\ Hesse, J.\ Roland, and D.\ Baye, 
                                 Nucl.\ Phys.\ {\bf A709}, 184 (2002).

\bibitem{Li6_SRG} E. D. Jurgenson, P. Navr\'atil and R. J. Furnstahl, arXiv: 1011:4085 [nucl-th].

\bibitem{cluster} P. Navr\'atil, Phys. Rev. C {\bf 70}, 054324 (2004).

\bibitem{Gruebler75} W. Gruebler, P. A. Schmelzbach, V. Konig, R. Risler and D. Boerma, Nucl. Phys. A, {\bf 242}, 265 (1975).

\bibitem{Jenny83} B. Jenny, W. Gruebler, V. Konig, P. A. Schmelzbach and C. Schweizer, Nucl. Phys. A, {\bf  397},  61 (1983).

\bibitem{Senhouse64} L. S. Senhouse, Jr. and T. A. Tombrello, Nucl. Phys. {\bf 57}, 624 (1964).

\bibitem{Jett77} J. H. Jett, J. L. Detch, Jr., and N. Jarmie, Phys. Rev. C {\bf 3}, 1769 (1971). 

\bibitem{NCSM_review} P. Navratil, S. Quaglioni, I. Stetcu and B. R. Barrett, J. Phys. G: Nucl. Part. Phys. {\bf 36}, 083101 (2009).

\bibitem{NCSM_RGM_INPC10} P. Navr\'atil, S. Quaglioni and R. Roth, arXiv:1009.3965 [nucl-th].

\bibitem{IT-NCSM} R. Roth and P. Navr\'atil, Phys. Rev. Lett. {\bf 99}, 092501 (2007).

\bibitem{Roth09} R. Roth, Phys. Rev. C {\bf 79}, 064324 (2009).

\bibitem{Varshalovich} D.\ A.\ Varshalovich,  A.\ N.\ Moskalev, and V.\ K.\ Khersonskij, {\em Quantum Theory of Angular Momentum} (World Scientific, Singapore, 1988).

\end{thebibliography}
\end{document}